\documentclass[preprint]{aastex}
\shorttitle{Dwarf galaxies in the Dorado group}
\shortauthors{Carrasco et al.}
\slugcomment{To appear in The Astronomical Journal, January 2001}

\newcommand{\Sec}{${}^{\prime\prime}$\llap{.}}
\newcommand{\Deg}{${}^{\circ}$\llap{.}}

\begin{document}

\title{The Dwarf Galaxy Population of the Dorado group down to $M_{V}\approx-11$\altaffilmark{1}}
\altaffiltext{1}{Based on the data collected at the Cerro Tololo 
Interamerican Observatory and  Las Campanas Observatory, Chile}

\author{Eleazar R. Carrasco and Cl\'audia M. de Oliveira}
\affil{Instituto Astron\^omico e Geof\'{\i}sico,Universidade de S\~ao
 Paulo\\ Caixa Postal 3386, 01060-970, S\~ao Paulo, Brazil}
\email{rcarrasc@pushkin.iagusp.usp.br,oliveira@iagusp.usp.br}

\and 

\author{Leopoldo Infante\altaffilmark{2,3}}
\affil{Dep. de Astronom\'{\i}a y Astrof\'{\i}sica, Facultad de
F\'{\i}sica\\ Pontificia Universidad Cat\'olica de Chile, Casilla 306,
Santiago 22, Chile}
\email{linfante@astro.puc.cl}
\altaffiltext{2}{Visiting astronomer, Cerro Tololo Interamerican
Observatory  (CTIO). CTIO is operated by Association of Universities for
Research in Astronomy Inc. (AURA),  under a cooperative agreement with the
National Science Foundation.}
\altaffiltext{3}{Visiting astronomer, Las Campanas Observatory. Las
Campanas is operated by the Carnegie Institution of Washington.}

\and

\author{Michael Bolte}
\affil{UCO/Lick Observatory, University of California, Santa Cruz, CA
95064}
\email{bolte@ucolick.org}

\newpage

\begin{abstract}

We present $V$ and $I$ CCD photometry of suspected low-surface brightness
dwarf galaxies detected in a survey covering $\sim2.4$ deg$^{2}$ around
the  central region of the Dorado group of galaxies. The low-surface
brightness galaxies  were chosen  based on their sizes and magnitudes
at the limiting isophote of 26.0V$\mu$. The  selected galaxies have
magnitudes brighter than $V\approx20$ ($M_{V}\approx-11$ for  an assumed
distance to the group of 17.2 Mpc), with central  surface  brightnesses
$\mu_{0}>22.5$ V mag/arcsec$^{2}$, scale lengths $h>2$\arcsec, and
diameters $\ge14$\arcsec at the limiting isophote. Using these criteria,
we identified 69 dwarf galaxy candidates. Four of them are large very
low-surface  brightness galaxies that were detected on a smoothed image,
after masking high  surface brightness objects. Monte Carlo simulations
performed to estimate  completeness, photometric uncertainties and
to evaluate our ability to detect  extended low-surface brightness
galaxies show that the completeness fraction is, on  average, $>80$\%
for dwarf galaxies with $-17<M_{V}<-10.5$ and  $22.5<\mu_{0}<25.5$
V mag/arcsec$^{2}$, for the range of sizes considered by us ($D\ge$
14\arcsec). The  $V-I$ colors of the dwarf candidates vary from -0.3 to
2.3 with a peak on $V-I=0.98$,  suggesting a range of different stellar
populations in these galaxies. The projected surface density of the
dwarf galaxies shows a concentration towards the group center similar in
extent to that found around five X-ray groups and the elliptical  galaxy
NGC1132 studied by Mulchaey \& Zabludoff (1999), suggesting that the
dwarf galaxies in Dorado are probably physically associated with the
overall potential  well of the group.  
\end{abstract}

\keywords{galaxies: clusters: general -- galaxies: photometry --
galaxies:  fundamental parameters (classification, colors, surface
brightness) --  galaxies: dwarfs: -- galaxies: luminosity function --
galaxies: cluster:  individual (Dorado group)}

\section{Introduction}

Dwarf galaxies are the most common type of galaxies in the local universe.
In addition, they are thought to be the single systems with the largest
dark-matter contents, with M/L ratios as high as that of groups and poor
clusters (e.g. Carignan and Freeman 1988) and hence their spatial
distribution and mass spectrum may give us important insights into the
spatial scales over which mass is distributed. However, they are also the
hardest galaxies to observe, due to their low-surface brightnesses and low
luminosities. Only recently it has become possible to obtain deep
wide-field CCD {\em photometry} of significant numbers of these galaxies,
mostly in clusters. {\em Spectroscopy} of such low luminosity systems is
still a challenge even with large-class telescopes, unless emission lines
are present.

Dwarf galaxies in nearby clusters and groups have been studied in detail
with photographic material (e.g.  Coma: Thompsom \& Gregory 1993; Virgo:
Sandage, Binggeli and Tammann 1985, Impey et al. 1988; Fornax: Ferguson
1989; Bothun et al. 1991; groups: Ferguson and Sandage 1991). From
these works two general trends were uncovered: 1) clusters and rich
groups usually have a luminosity function with a faint end that is
steeper, i.e., has more low-luminosity galaxies, than that for poorer
groups and 2) Dwarf elliptical and dwarf spheroidal galaxies (dE/dS0)
seem to dominate the faint-end of the number counts in clusters while
the dwarf-irregular population (dIrr) are more representative in poor
groups and in the field.  Point (1) was well illustrated by Ferguson \&
Sandage (1991) who found a faint-end slope (fit to a Schecter function)
of $\alpha=-1.3$ in the composite luminosity function of seven nearby
groups (the sample included the Dorado group) with a tendency for the
richer groups to present a larger ratio of dwarf to giant galaxies than
the poorer groups.  Point (2) has been challenged by observations of the
most well known group of all, our own Local Group. Despite being a poor
group with just over 30 members, the Local Group has a larger number of
dE/dS0 than of dIrr (Pritchet \& van den Bergh 1999).  The majority of
the dwarf spheroidal galaxies are concentrated around M31 and the Milky
Way with a slope of the faint-end of the luminosity function (fit to a
Schechter function) of $\sim -1.3$ while the dwarf irregulars are less
concentrated with a flat luminosity distribution (van den Bergh 1999,
Grebel 2000). This seems to contradict the results  obtained for the
CfA groups and field survey where the largest contribution to the steep
faint-end was from the irregular galaxies (Marzke et al. 1998). Recent
results coming from CCD studies in nearby groups up to $D=10$ Mpc show
an increasing number of new dwarf members (e.g. Jerjen et al. 2000)
with a spatial distribution similar to that found in the Local Group
(i.e the dwarf spheroidals concentrated around the brightest galaxies
while the irregulars distributed throughout the group).

This is the first paper of a series dedicated to survey nearby groups
with  the aim of identifying low-surface brightness dwarf galaxies
(hereafter LSBD)  down to $M_{V}\sim -11$. Our primary goal is to study
the physical properties  and luminosity distributions of the LSBD and
ultimately determine the  luminosity function of the groups as a function
of their richness and other  group parameters.  Here we report the results
of the CCD photometry of the dwarf galaxy candidates of the nearby group
of galaxies  Dorado. A further study will analyze other nearby groups
of galaxies.

The Dorado group is  a loose concentration of galaxies centered at $\sim
4^{h} 15^{m}$,  $-55^{\circ} 41{'}$ (epoch J2000).  It is also known as
Shk 18 (Shakhbazian 1957),  G16 (de Vaucouleurs 1975) and HG3 (Huchra
\& Geller 1982).  Dorado was first identified by de Vaucouleurs (1975)
as a ``large and complex nebulae'' in the region of Dorado. In 1982,
Huchra \& Geller (1982), measured velocities for 18 galaxies in the
group. Maia et al. (1989) extended the number of known members to 46.
Finally Ferguson \& Sandage (1990, 1991), in a study of the galaxy
properties in seven nearby groups identified 34 possible members of
Dorado with magnitudes $B_{T} > 19$ in photographic plates. The central
part of the group is dominated by two almost equally bright early-type
systems: NGC1549 (an E1 galaxy with $V_{r}=1247$ km/s) and NGC1553 (an
S0 galaxy with $V_{r}=1280$ km/s).  No detailed distance determination
has been made for Dorado and therefore we will adopt a group distance
of 17.2 h$^{-1}$ Mpc, where $h=H_{0}/75$. This was determined for the
central galaxy NGC1549 based on the D$_n$-$\sigma$ relation from Faber
et al. (1989).  For the determination of the absolute magnitudes of the
galaxies we chose the value of $H_{0}=75$ km s$^{-1}$ Mpc$^{-1}$. The
group covers an area on the sky of $\sim$ 10\arcdeg $\times$ 10\arcdeg
($\sim$ 3$\times$3 h$^{-1}$ Mpc).

This paper is arranged as follows. Section 2 describes the observations,
the data reduction and the methodology used to detect, classify and to
obtain photometric parameters for galaxies in the frames. In that section
we  also describe the Monte-Carlo simulations performed in order to
estimate  the completeness for the LSBD population, and to determine the
distribution  of detected central surface brightness and scale lengths to
be expected if Local Group and Virgo-like LSBD were present in the Dorado
group. In section 3 we present the results and analyze the central surface
brightness, scale length, spatial and  color distributions of the LSBD
candidates. Finally, in section 4, a general  discussion and a brief
summary of the main results are presented.  The luminosity function of the
group will be determined, along with those for other groups, in a later
publication.

\section{Observations and data reduction}

\subsection{Observations}

The Dorado group was observed in two separate runs at the end of
1996. The central region of the group was observed on November 1--2,
with the 0.9m Curtis/Schmidt telescope at the Cerro Tololo Interamerican
Observatory (hereafter CTIO). The images were obtained using a standard
STIS  $2048 \times 2048$ pixel CCD with 21 $\micron$ pixels corresponding
to 2.028$\arcsec$/pixel on the sky. The total field of view was 1\Deg15
$\times$ 1\Deg15.   The second set of observations was  performed with the
1.0m Swope telescope  at the Las Campanas Observatory  (hereafter LCO40),
Chile, in the nights of  December 2-7, 1996. In this case the Tektronix 5
CCD, with $2048 \times 2048$ 21-$\micron$ pixels (0\Sec696 on the sky) was
used. The total field of view per image was 24\arcmin $\times$ 24\arcmin.

The fields were imaged through standard Johnson $V$ and Cousins $I$ filters. 
Typical total exposure times were between 30 and 60 min per field in both
filters. All data were taken under photometric conditions, with a seeing
ranging from 1\Sec3 to 1\Sec9 in the LCO40 frames. Due to the large pixel
size of the CCD used in the observations taken at CTIO (2.028\arcsec/pix),
the image quality in the central field is much worse than that for the LCO40 
images ($\sim$3\arcsec). Table 1 shows the  observation log (field
identification,  equatorial coordinates, filters, observatories, exposure
times, the observed area and the area used on the analysis after the masked
pixels were  disregarded) where  D01 is the central field observed in CTIO
and D02 to D11 are the fields  observed in LCO40.  The total area covered by
the observations was $\sim$ 2\Deg6 ($\sim$ 1\Deg2 in the central region and
$\sim$ 1\Deg4 spread over the  outskirts of the group) in each of the
filters. After subtraction and masking the bright objects in the fields, the
total area used in the analysis was $\sim 2$\Deg4. Figure \ref{fig1} shows
the sizes and locations  of the observed CCD images. The symbols indicate the
galaxies with known velocities taken from the NED. In addition, we observed
two fields 10\arcdeg  north (B3) and 15\arcdeg south (B4) of the Fornax
galaxy NGC 1399, in order  to estimate the background contamination of low
surface brightness galaxies. These observations  were performed in the second
run (LCO40), with the same filters and exposure times matching  the Dorado
LCO40 observations  (see Table 1 for details). 

\subsection{Reductions and photometric calibrations}

The images were bias/overscan-subtracted and trimmed, using the CCDPROC task
inside IRAF\footnote[1]{IRAF is distributed by NOAO, which is operated by 
the Association of Universities for Research in Astronomy Inc., under contract
with the National Science Foundation}. Dome and twilight flats were combined
with the object-frames taken during the night to make a master flat-field and
used to flat-field the data. The images were then combined by obtaining their
median. The final combined images had an average {\em rms} pixel-to-pixel
variation of the sky level that corresponded to a limiting surface brightness
of $\sim 26.2$  (CTIO) and $\sim 26.8$ $V$ mag/arcsec$^{2}$  (LCO40) for the
$V$ filter and $\sim 25.2$ (CTIO) and $\sim 26$ (LCO40) for  the I  filter.
In the case of the LCO40 I frames, there were small scale  variations along
the CCD and since these could affect the analysis of the  structural
parameters of the LSBD, we decided to do the photometry on the I images using
the V images as templates (i.e. using the same positions, ellipticities and
areas obtained for objects in the V filter). The main use of the I photometry
was to determine aperture colors for the objects.

Calibrations of the magnitudes to the standard system were derived
using observations of standard stars from Landolt (1992). A number of
standard stars were observed throughout the nights for that purpose: 26
and 22 during the two nights for CTIO and 42, 46, 47, 47  and 44 in the
five nights for LCO40. The task APPHOT in IRAF  was used to determine
the instrumental magnitudes of the stars.  Transformation equations
from the instrumental to the standard Johnson-Kron-Cousins system were
obtained using the following relation:

\begin{displaymath}
V = v+a^{v}_{0}+a^{v}_{1} X_{V}+a^{v}_{2} (V-I)
\end{displaymath}
\begin{equation}
I = i+a^{i}_{0}+a^{i}_{1} X_{I}+a^{i}_{2} (V-I)
\label{equ1}
\end{equation}

\begin{displaymath}
(V-I) =  \frac{ [(v+a^{v}_{0})-(i+a^{i}_{0})]+a^{i}_{1}X_{I}-a^{v}_{1}X_{V} }{1-(a^{v}_{2}-a^{i}_{2})}
\end{displaymath}

\noindent where {\em v} and {\em i} are the instrumental aperture magnitudes 
of the standard stars, $a^{v}_{0}$ and $a^{i}_{0}$ are the transformation
coefficients between the standard and instrumental system, $a^{v}_{1}$ and
$a^{i}_{1}$ are the extinction coefficients, and $a^{v}_{2}$ and $a^{i}_{2}$
represent the coefficients of the color term. The calibration equations and
coefficients for both CTIO and LCO40 runs are given in Table 2 (col. 4). The
{\em rms} of  the difference between standard and calibrated magnitudes for
the standard  stars is less than 0.025 mag (col. 5 in Table 2). We can see in
Table 2 that the color coefficients are small and stable during the nights
(LCO40 observations). The zero point for each filter was obtained using the
two first equations of (1) and the coefficients given in Table 2 with an
assumed color index of $(V-I)=1$, a typical value for the LSBD 
population in nearby groups and clusters like Virgo (Impey \& Bothun 1997). 
Magnitudes and colors were then calibrated accordingly.

\subsection{Detection and photometry of the objects}

Detection, photometry and classification of objects were performed using the
Source Extractor (SExtractor) software program (Bertin \& Arnouts 1996).
Before running SExtractor, we removed all bright galaxies and saturated
stars. These are necessary steps since otherwise the bright objects may have
either prevented detection of faint sources or significantly affected the
photometry of objects located  in their wings. Figure \ref{fig2} shows the
central region of the Dorado group observed at CTIO (upper panel) with some 
examples of low surface brightness dwarf galaxies detected in this region
(see below). The four brightest galaxies in the field,  NGC1549, NGC1553,
NGC1546 and IC2058 are marked on the figure (upper panel). These four
galaxies were extracted from the frames, first by fitting their isophotes
using the task STSDAS.ISOPHOTES.ELLIPSE.  Then, their fitted ellipses were
subtracted from the frame using the task BMODEL. Objects  which were poor
fits to elliptical isophotes were simply masked.  Masking was also used to
remove bright saturated stars and the remaining residuals from the bright
elliptical galaxies  subtraction.  The same procedures were applied to
subtract or mask the bright galaxies and stars from the LCO40 frames. In
these frames only one galaxy was  subtracted using the fitting technique
(NGC1536 in D05). These pixels, corresponding to the masked areas, were not
used in subsequent analysis.

The analysis of the image with SExtractor proceeded as follow: {\it i)} in
each CCD image the background map was constructed using  sub-areas of $64
\times  64$ pixels ($\sim 130 \times 130$ arcsec) and $128 \times 128$ pixels
($\sim  90 \times 90$ arcsec) for CTIO and LCO40 images respectively. These
values  are much larger than the typical size of a low-surface brightness
galaxy that could be  detected at the distance of the group (searches for
larger galaxies were conducted in Section 2.7); {\it ii)} the local
background  histogram was clipped iteratively until convergence at $3\sigma$
was reached.  These raw background values were filtered using a median filter
(in 3 $\times$  3 sub-areas), and a bi-cubic-spline was used to obtain the
resulting  background map. This background map was then subtracted from the
image;  {\it iii)} the images were convolved with a {\em top-hat} mask with a
FWHM  slightly larger than the values for the stellar images, in order to
enhance  the efficiency in detecting low-surface brightness, extended
objects. All  objects with a threshold $\ge$ 26 V mag/arcsec$^{2}$ ($\sim$
1.2$\sigma_{sky}$ and $\sim$ 1.7$\sigma_{sky}$ for the CTIO and LCO40 frames
respectively) above the  sky level and with a minimum area of 5 pix$^{2}$ 
were found and their  photometric parameters obtained (note that a threshold
of $\sim$  1.2$\sigma_{sky}$ for the LCO40 images correspond to a limiting
surface  brightness of $\sim$ 26.6 V mag/arcsec$^{2}$, i.e. 0.6 V
mag/arcsec$^{2}$  fainter than our limiting surface brightness).  We used a
threshold of 26 V mag/arcsec$^{2}$ in order to have a common surface
brightness cut in both the CTIO and the LCO40 fields. 
  
Photometry of all objects in the $V$ and $I$ frames was done using elliptical
apertures. Total magnitudes were computed using the Kron's ``first moment''
algorithm $r_{1}=\sum r I(r)/\sum I(r)$ (Kron 1980, Infante 1987). For each
object, the elliptical apertures are chosen and the ellipticity and the 
position angle are defined by the second-order moments of the light 
distribution. The maximum aperture is then defined as one that has double
the  area of the isophotal elliptical aperture. Inside this aperture, the
integrated  light is used to construct a growth curve where the first moment
is computed. As discussed by Infante (1987) and Infante \& Pritchet (1992), 
the $2r_{1}$ measures about 90\% of the flux for galaxies with  exponential
profiles, and about 96\% for objects with stellar profiles. We checked these
values with  Monte Carlo simulations (see section 2.6). It may also
underestimate the true  total magnitude for galaxies with $r^{1/4}$ intensity
profiles, as suggested by Secker \& Harris (1997. However, the number of
faint group galaxies  with $r^{1/4}$ is small.

The object detection was performed in the V filter and was used as a
template  to obtain the photometric parameters in the I filter (i.e. we
forced the  detection in the I filter of those objects detected in the V
filter using  common structural parameters).  The output catalogues in both
filters were then matched  in order to obtain color information for all
objects. Colors were determined by measuring magnitudes in a circular
aperture of 7\arcsec~ of diameter in the  CTIO field and 4\arcsec~ in the
LCO40 fields, in both filters. Those objects  centered within $\sim
10$\arcsec of any edge or masked region in the images  were deleted from the
catalogues. The output matched catalogues (available  under request) contain
the isophotal  magnitude, the isophotal area, the  total magnitude, the 
ellipticity, eight isophotal areas which correspond to  different surface
brightnesses (see section 2.5), the peak surface brightness  in
mag/arcsec$^{2}$ (this corresponds to the intensity at the central pixel), 
coordinates at epoch J2000.0, a quality flag and the star/galaxy
classification for each object.

The completeness of our sample for point sources falls below 90\% at 
magnitudes $V\sim20.5$ mag and $V\sim23$ mag for the CTIO and LCO40 
observations respectively (see section 2.6).  For the $I$ fields, the 
completeness falls to 90\% at magnitudes $I\sim19$ mag and $I\sim21$  
mag for the CTIO and LCO40 observations  respectively.

\subsection{Star/Galaxy separation}

The star/galaxy separation (resolved $=$ galaxies, unresolved $=$ stars,
globular clusters, unresolved background galaxies) was done with a neural
network routine developed by Bertin \& Arnouts (1996). This is a training
program that classifies objects with an index (stellarity index)  between 0
(galaxies) and 1 (stars). It uses eight isophotal areas (see section 2.5),
the peak intensity, and the seeing  to classify the objects. The star/galaxy
separation works well down to $V\sim20$ (CTIO) and $V\sim22$ (LCO40). 
Objects with magnitudes beyond this limit present an increasing scatter in
the classification.
 
In order to check the star/galaxy classification given by SExtractor we first
did a visual inspection of all objects classified as galaxies with
``stellarity index'' less or equal than 0.3. The choice of 0.3 is based on
our Monte Carlo simulations (section 2.6) since galaxies that mimic Local
Group  LSBD at the distance of Dorado are classified in all cases with
stellarity indices lower than 0.3.  Eye control showed that the
classification of galaxies is adequate down to  $V\sim20.0$ and $V\sim22.0$
for the CTIO and LCO40 respectively. For fainter magnitudes, the
classification is uncertain. An alternative method was used to check the
classification. This method consisted in plotting pairs of parameters like 
the central intensity {\it vs.} area,  the central intensity {\it vs.} size
of the objects, and  the peak intensity {\it vs.} size of the objects. For
each plot, a curve was drawn in order to define the boundaries between  stars
and galaxies. Then, the classifications obtained from each of the three plots
were combined to  assign to each object a probability that it is a star, a
galaxy or noise.  We found that $\ga 92$\% of the galaxies were classified
identically by this method, by SExtractor (``stellarity index'' $\le0.3$) and
by the visual inspection down to $V\sim20$ and $V\sim22$ mag for the CTIO and
the LCO40 frames respectively. The remaining objects were difficult to
classify and these were labeled with an ``uncertain'' flag. 

\subsection{Selection of the low-surface brightness dwarf galaxy candidates}
 
Selection of the LSBD candidates was done using parameters given by the
exponential profile fit (the central surface brightness and scale length),
and the diameter at the limiting isophote. Below we describe in detail our 
procedure.

The radial profile of the galaxies can be represented by a generalized 
exponential law or Sersic law of the form:

\begin{equation}
\mu(r) = \mu_{0} + 1.086 (r/h)^{n}, \quad n>0
\label{equ3}
\end{equation}
 
\noindent where $\mu_{0}$ is the central surface brightness, $h$ is the
scale factor and $n$ is the generalized exponent. The surface brightness
profile of disk galaxies and dwarf galaxies fainter than $M_{V}\sim-17$ are,
in general, well fit by a pure exponential law, with $n=1$ in (\ref{equ3}) 
(Faber \& Lin 1993, Binggeli \& Cameron 1991, 1993).

The fit to a pure exponential law was used to obtain the extrapolated central
surface brightness and the scale length of the LSBD candidates in our sample.
For each of the eight isophotal areas given by SExtractor we obtained  the
surface brightness and the radius using a relation of the form

\begin{equation}
\mu_{i}= \frac{i}{8}\mu_{peak}+\left(1-\frac{i}{8}\right) \mu_{lim} \quad
\quad \quad i = 0,...,7
\label{equ4}
\end{equation}

\noindent where $\mu_{peak}$ is the surface brightness at the pixel of
maximum intensity, $\mu_{lim}$ is the surface brightness at the limiting
isophote, and $\mu_{i}$ is the surface brightness at the given isophotal
area. Each value of the surface brightness corresponds to a radius
$r_{i}=\sqrt{A_{i}\epsilon/\pi}$, where $\epsilon=a/b$ is the elongation and
$A_{i}$ is the isophotal area at the {\em i}-label given by  SExtractor.
Using the values for $\mu_{i}$ from (\ref{equ4}), we made a least-squares fit
to obtain the extrapolated  central surface brightness and the scale lengths
for each galaxy. 

We used the extrapolated central surface brightness and scale length given by
the fit as the main parameters to allow a primary cut on the catalogue. All
galaxies with central surface brightnesses fainter than $22.5$ V
mag/arcsec$^{2}$ and scale lengths h$>$3\Sec0 and h$>2$\Sec0 for the CTIO and
LCO40 observations respectively were selected (note that the seeing strongly
affects these structural parameters for $h<3^{''}$ and 2$^{''}$
respectively).  This surface brightness cut is similar to that used by
Dalcanton et al. (1997) to study the number counts of LSBD in the local 
universe and is one magnitude fainter that the value used by O'Neil et al. 
(1997) to search for low-surface brightness galaxies in the Pegasus and
Cancer  clusters. The first cut on the data selected 777 objects in all
fields.

The second cut was done selecting galaxies with a limiting diameter 
$D_{lim}=2\sqrt{A_{lim}\epsilon/\pi}\ge14$\arcsec at the limiting surface 
brightness of 26 V mag/arcsec$^2$. The diameter at the limiting isophote 
(limiting diameter hereafter) can  be expressed also by the parameters given
by the exponential fit (Allen \& Shu 1989):

\begin{equation}
\theta_{lim}=0.735 (\mu_{lim} - \mu_{0}) 10^{0.2(\mu_{0}-m_{tot})}
\label{equ5}
\end{equation}

\noindent where $\mu_{lim}$ is the surface brightness at the limiting
isophote of 26 V mag/arcsec$^2$,  $\mu_{0}$ is the extrapolated central
surface brightness, and $m_{tot}$ is the total magnitude of the objects. In
our case, a diameter of  $\ga 14$\arcsec~ represents well the expected sizes
of LSBD that can be  detected.  This is supported by Monte-Carlo simulations
(see next section),  which show that the fraction of detected LSBD with
exponential profiles with  $22.5 \le \mu_{0} \le 24.5$ ($22.5 \le \mu_{0} \le
25.5$ for LCO40 fields)  and diameters $\ge 14''$ is $f\ga80$\% down to
$V\sim20.0$ ($V\sim22$). The second  cut on the data (size limit) limited the
sample to 144 LSBD candidates. Each of these galaxies were inspected visually
and 84 of them were discarded. We eliminated these galaxies based on the {\em
flag} of the photometric quality (Bertin \& Arnouts 1996). We found that all
objects with {\em flag} $>0$ have a bright object in their proximity (with
the additional  contaminating light from the nearby bright source, the false
LSBD's were  artificially included in our catalogue). This last cut limited
the sample to 60 LSBD candidates. The same selection criteria were used to
find LSBD in the two  control fields (LCO40) in order to estimate the
contribution of background  galaxies to the luminosity function. We found
only one galaxy in both control  fields  with $\mu_{0}\ge22.5$ V
mag/arcsec$^{2}$, diameter $\ge14$\arcsec, scale  length $h>2$\arcsec. This
galaxy seems to be a genuine low-surface brightness galaxy in the field with
apparent magnitude $V=18$ and color $V-I=1.1$.  In addition to the 60 LSBD
galaxies, we found five LSBD galaxies in the LCO40  fields with
$\mu_{0}\ge22.5$ V mag/arcsec$^{2}$, scale length $h>2$\arcsec and diameter
$\ge14$\arcsec at the limiting surface brightness of $\mu_{lim}\ge26.5$  V
mag/arcsec$^{2}$ (see section 2.7.1).

Table 3 lists the photometric data and the profile fitting parameters of  the
LSBD detected in the group (the last four lines of the  table  correspond to
the large LSBD namely L01 to L04 - see section 2.7). The  data are given as
follows: column 1 provides the galaxy identification;  columns 2 and 3 list
the equatorial coordinates at the epoch 2000.0;  columns 4 and 5 show the
``total'' magnitude (SExtractor ``auto'' magnitude,  not corrected according
to the simulations and not corrected by  interstellar absorption) in $V$ and
the $V - I$ colors within apertures of  7\arcsec and 4\arcsec in diameter for
CTIO and LCO40 data respectively; columns  6 and 7 list the  limiting
isophotal diameter and the diameter given by relation (\ref{equ5})  at the
surface brightness of 26 V mag/arcsec$^{2}$; the extrapolated  central
surface brightness $\mu_{0}(V)$ and the scale lengths are  given in columns 8
and 9; the next two columns (10 and 11) show the  effective radius and mean
effective surface brightness; the last column  shows the morphological type
(see section 3.1). The effective radius in column  10, defined as the radius
of the  isophote that contains half of the  total light, can be represented,
in the case of a pure exponential law, by  $r_{eff}=1.668h$ and the mean
effective surface brightness (column 11) by 
$\mu_{eff}=\mu(r_{eff})=\mu_{0}+1.086(r_{eff}/h)$ (Binggeli \& Cameron 1991).
The equatorial coordinates in Table 3 (columns 2 and 3) were determined 
using 20-30 stars from the Guide Star Catalogue present on the images  giving
an accuracy in the position of less than 2\arcsec. 

\subsection{Photometric uncertainties and completeness of the sample via Monte 
Carlo simulations}

We performed Monte Carlo simulations in order to determine:  {\it i)}
the completeness for point sources; {\it ii)} the completeness for LSBD
galaxies (i.e. the fraction of them detected with the automatic
detection described in Section 2.3); and {\it iii)} the uncertainties in
the photometry of the LSBD galaxies.  The simulations were performed on
the science frames in order to have the same characteristics of a real
image (cosmetic defects, crowded images, light gradients, noise and
seeing).

We first computed the completeness of the sample by adding point
sources constructed from the observed point spread functions and
checking the fraction of them we could recover. The results are shown
in Figure \ref{fig3} and quantified in Table 4. From the figure we can
see that our completeness limit for point sources with ``stellarity
index'' $\ge 0.9$ is $\ge 90$\% to the limits of $\sim 20$ and $\sim
23$ for the CTIO and LCO40 frames respectively.

We then computed the completeness of the sample for galaxies with
magnitudes, central surface brightnesses and scale lengths typical of
low-surface brightness dwarf galaxies  at the redshift of the group. For
this we assumed that the LSBD have  an exponential profile. For each
image, we simulated galaxies with exponential profiles with different
scale lengths, central surface brightnesses and random ellipticities. Each
image was convolved with a point spread function constructed from bright
stars in the frame.  The method implemented was the following: {\it i)}
for each bin of central surface brightness we generated 400 randomly
distributed disk galaxies in groups of 20 (five for the brightest  bins)
with magnitudes between 16.0 -- 20.0 and 16.0 - 22.0 for the CTIO
and LCO40 frames respectively.  These galaxies were created using the
MKOBJECTS program in the NOAO.ARTDATA package in IRAF; {\it ii)} we ran
the  SExtractor program with the same detection parameters previously
used for the real-galaxy detection (see section 2.3).  Common objects
in the input and output catalogues were matched; {\it iii)} using the
eight isophotal areas, for each matched galaxy given by SExtractor,
we calculated the central surface brightness, the scale length and the
limiting diameter using equations (\ref{equ3}) (for $n=1$), (\ref{equ4})
and (\ref{equ5}) of section 2.5;  {\it iv}) finally,  the photometric
and structural parameters were compared  with the input parameters.

Table 5 lists the average completeness fraction for LSBD with
$22.5<\mu_{0}<25.5$ mag/arcsec$^{2}$ ($22.5<\mu_{0}<26.5$ for the
LCO40 frames), $16<V_{tot}<21$ and $\theta_{lim}$ $>$ 14\arcsec. The
fractional values  are the averages over the ranges of surface brightness
(per 0.5  bin in magnitude). Figure \ref{fig4} shows the same results
in graphical  form.  The completeness is, on average, $f\sim80$\% in all
cases except  for the last bins in surface brightness ($24.5<\mu_{0}<25.5$
and  $25.5<\mu_{0}<26.5$ for CTIO and LCO40 frames respectively). In those
cases the completeness fraction is less than 40\%. The completeness in
magnitude falls abruptly for $V>20$. Table 6 shows the average differences
between input and output magnitudes, central surface brightnesses
and scale lengths (in percentage of $(h_{inp}-h_{out})/h_{inp}$).
Figure \ref{fig5} shows differences between input and output magnitudes
in a  graphical form. The differences in magnitude are lower than 0.3 mag
(solid symbols in both panel in Fig.  \ref{fig5}) for $\mu_{0} < 24.5$
(CTIO)  and $\mu_{0} < 25.5$ (LCO40).  In this interval, the average
differences between input and output central surface brightnesses are
less than 0.5 V mag/arcsec$^{2}$. For the scale lengths, our simulations
show that output values are, on average, larger than input ones by
$\sim20$\%. For the last surface brightnesses bin in the CTIO and LCO40
data, these differences are $\sim$40\% and $\sim$30\% respectively (last
two columns in Table 6). This result shows our limitation in estimating
in a correct way the scale length of our real galaxies especially at
small sizes. The values for the scale length listed in Table 3 with $h <
5$\arcsec and $M_{V} > -12.5$ are only upper limits (see also Fig. 8a
in section 3.2). In the next section we describe an  alternative method
used to detect large galaxies with very low surface brightnesses. It
increases  the completeness fraction in the last bin of central surface
brightness to $\sim80$\%. No galaxies with magnitudes brighter than
18 and  central surface brightnesses $\mu_{0}>24.5$ were found in our
sample. These  objects would have scale lengths $h>8$\arcsec. These have
not been found in  other surveys either.

\subsection {Searching for large low-surface brightness galaxies}

\subsubsection {Galaxies below the $\mu_{lim}=26$ in the LCO40 fields}

The limiting surface brightness imposed by the central field of
$\mu_{lim}=26$ represents 1.7$\sigma_{sky}$ in the LCO40 fields taken in
the outskirts of the group. We therefore made an additional search for
galaxies with diameters larger than 14'' but at limiting isophotal level
of 26.5 V mag/arcsec$^2$ (or 1.2$\sigma_{sky}$) only on the LCO40 fields.
In this way we discovered five new galaxies that are listed in Table 3
with a ``d'', where ``d'' stands for deep.  The inclusion or exclusion
of these galaxies in the analysis described below does not change the
results.

\subsubsection {Large low-surface brightness galaxies in all fields}

As described in section 2.3, the sky subtraction was done by taking
the median values for image sections of $64 \times 64$ pixels and $128
\times 128$ pixels in size.  This limiting size, used for making the
background map could erase any signs of larger galaxies in our frames,
if they were present. We then repeated the reduction procedure using
background sub-sections of 128 and 256 and re-did searches for galaxies
with larger sizes. This was done by smoothing the images after masking all
high-surface brightness objects. The detection was done in the smoothed
images and the photometric and structural parameters were obtained in
the unconvolved images. The advantage of the smoothing technique used
to reduce the pixel-to-pixel noise is that it may be done with a filter
that is matched to the shape and size of the galaxy which we are looking
for; this produces the maximum gain in signal-to-noise (enhancement
of the image) for the targeted galaxies. The original data can provide
additional information on any low-surface brightness features detected
after smoothing.

The procedure used was the following: {\it i}) all objects with 3$\sigma$
above sky were masked and substituted by the appropriate sky noise of
each image; {\it ii)} an additional area around each object in ({\it
i}) was also masked in  the same way, to ensure that all high-surface
brightness areas of the image were masked.  A 2-pixel ``growing radius''
was used for this purpose (i.e., the radius of each object in ({\it i})
was incremented by 2 pixel for masking); {\it iii}) the ``clean'' image
from ({\it ii}) was convolved with gaussian functions of different widths
and for each resulting image we ran SExtractor in order to look for large
remaining low-surface brightness objects; {\it iv)} the photometry of
the objects detected in the smoothing images were done in the original
frames using the convolved images as templates. At this point, we also
did an eye inspection of the images to check if the automatic search
with SExtractor had lost any objects.

In this way a number of new very low-surface brightness objects were
uncovered in the CTIO frame (four galaxies). No additional galaxies
were found in the LCO40 frames at 1$\sigma$ sky level. The last lines of
Table 3 show  the photometric and structural parameters for these four
galaxies. No color information is available since these were not detected
in I. The three panels in Figure \ref{fig6} exemplify our procedure. The
upper-left panel shows the masked image with an extended, low surface
brightness galaxy that could not be detected by SExtractor using the
procedure in section 2.3 (galaxy L01 in Table \ref{tab3}). The upper-right
panel shows the same object after smoothing by a gaussian filter. Finally,
the lower panel shows the SExtractor detection on the smoothed image.

One of the detected low-surface brightness object (L02 in Table
\ref{tab3}) is very extended and has an irregular morphology and very
low-surface brightness.  This ``cloud'' has a size of $\sim$ 2\arcmin
$\times$ 1\arcmin (10 $\times$ 5 h$^{-1}$ kpc), much larger than the
value obtained with SExtractor (see Table \ref{tab3}). This ``cloud'' is
located $\sim$ 10\arcmin NW of the central galaxy NGC1549. A discussion
about the nature of this object is given in section 4.

In order to quantify our ability to detect the extended LSBD objects
and to evaluate the photometric errors of this technique, we have
performed Monte Carlo  simulations in a similar way to what was
discussed in  section 2.6, but applying a  smoothing filter. For each
image, we applied the method explained above to obtain a smoothed
and cleaned frame and to detect extended, LSBD in the images with
SExtractor. Figure \ref{fig7} shows a simulated galaxy with $V\sim18.5$
and $25.5<\mu_{0}<26.5$. The results of the simulations show that the
completeness fractions  increase from 40\% (obtained with the simulation
in section 2.6) to $\sim 80$\% (with the technique employed here)  for
galaxies with $24.5<\mu_{0}<25.5$ (CTIO), $25.5<\mu_{0}<26.5$ (LCO40),
and  magnitudes between $18<V<20$. The differences in magnitude (input
-- output)  is $\sigma(\Delta m) < 0.3$ mag (open symbols in figures
\ref{fig4} and \ref{fig5} respectively). The results of these simulations
for the last two bins of surface brightness are quantified in Table 6.

\section{Results}

Here and in the following sections we will assume that all detected LSBD
galaxies in our sample are group members.

\subsection{Morphological types of the LSBD galaxies}

The LSBD images in the central region of Dorado are of too low resolution
to allow a good morphological classification. In this region only a few
of the largest galaxies can be classified. However, for the LCO40 data,
taken at the outskirts of the group, a classification of all galaxies
was attempted (including the 5 LCO40 galaxies with limiting surface
brightness 26.5 V mag/arcsec$^{2}$). Our visual classification of
27 galaxies (9 in the central region, including the large low-surface
brightness galaxies) shows that 6 are clumpy, which we classify as dwarf
irregular galaxies and 21 have smooth morphology, classified in a broad
way as dwarf elliptical galaxies. From the 21 dwarf elliptical galaxies,
three, and possibly another 3 in the central region, show bulge and disk
(see below). Three of the dwarf irregular galaxies in the LCO40 frames
seem to have star forming regions.

\subsection{Central surface brightness, scale length and color
distributions}

Figure \ref{fig8} shows $\log_{10} h - M_{V}$ and $\mu_{0} - M_{V}$ diagrams
for the 60 LSBD galaxies detected in the region of the Dorado group down
to $\mu_{lim}=26$ V mag/arcsec$^{2}$ after the 
cuts described in section 2.5. It also shows four additional large low-surface 
brightness galaxies found with a smoothing technique (see section 2.7) and
the five galaxies detected in the LCO40 fields down to $\mu_{lim}=26.5$  
V mag/arcsec$^{2}$ (section 2.7.1). The 
total magnitudes were corrected by the flux loss by SExtractor (aperture
corrections). These values were obtained in  the simulations and are 
quantified in Table 6, section 2.6. The magnitudes were also corrected for 
interstellar absorption. The absorption correction was obtained from the 
reddening maps of Schlegel et al. (1998), using the relations 
$A(V)=3.1 \times E(B-V)$ and $A(I)=1.5 \times E(B-V)$ taken from Cardelli
et al. (1989). At the position of the group, the absorption correction for the 
V filter is less than 0.04 mag and for the colors, the correction is less than 
0.02 mag. In Figure \ref{fig8} we plotted also, for comparison, the Local 
Group (LG) dSph galaxies with $M_{V}>-16$ (filled pentagons) redshifted to the 
distance of Dorado. The absolute magnitudes, central surface brightnesses and
scale lengths of the LG dSph galaxies were obtained from Lin \& Faber (1983),
Caldwell et al. (1992), Da Costa (1994) and Irwin \& Hatzidimitriou (1995). As 
we can see from the plots, several dSphs of the Local Group could be detected 
at the distance of Dorado, if they were present there. The faintest Local Group 
dSphs, with absolute magnitudes fainter than $M_{V}=-11$, like Draco, Carina, 
Sextans and Ursa Minor, would  not be detected at the distance of Dorado.

We find no clear correlation in the $\mu_{0}-M_{V}$ plane (Fig. 8b) and a
weak correlation is seen in the scale-length -- $M_{V}$ plane (Fig. 8a).
Correlations in these two planes do not have physical meaning since they
may be produced by selection effects. The night-sky brightness introduces
a strong selection effect and restricts the true determination of the
luminosity distribution of galaxies.  Both very compact, high-surface
brightness objects and extended, very low-surface brightness galaxies are
hard to detect.  However, the numbers of compact high-surface brightness
in groups may be small (see for example Drinkwater et al. 1996).

The faintest LSBD found in Dorado has an absolute magnitude of
$M_{V}=-11.1$ (V$\sim20.1$), with a central surface  brightness of
$\mu_{0}=22.9$ V mag/arcsec$^{2}$ and a scale length $h=2.6$\arcsec
($\sim 0.3$ h$^{-1}$ kpc). The largest LSBD galaxy ($D_{26}\sim69$\arcsec)
has a scale length of  $h=11.1$\arcsec (0.9 $h^{-1}$ kpc), $M_{V}=-15.5$
and $\mu_{0}=22.7$ V mag/arcsec$^{2}$. The colors inside a diameter of
7\arcsec (CTIO) and 4\arcsec (LCO40) of the 65 LSBD galaxies detected
in Dorado with a color information vary from $-0.3 < V-I < 2.3$ with a
peak at $V-I=0.98$ (see Figure \ref{fig9}).

We have obtained the surface brightness profiles of the 65 LSBD using
the task ELLIPSE in the STSDAS package inside IRAF.  We then used the
procedure described in section 2.5 to obtain the extrapolated central
surface brightness and scale lengths by a least squares fitting procedure
(we used here only the outer parts of the profiles to avoid the region
affected by seeing). The comparison between the results given from the
SExtractor areas (see section 2.5) and by ELLIPSE are similar (within
0.2 V mag/arcsec$^2$ for $\mu_{0}$ and within 20\% for the scale length).
These results are also in good agreement with the Monte Carlo simulations
(see section 2.6). Figure \ref{fig10} shows the surface brightness
profiles for the 65 LSBD galaxies (we do not include the four large
low-surface brightness galaxies listed at the end of Table \ref{tab3}
because of the poor fit to the data). In this figure the x-axis is
the semi-major axis of the elliptical isophote. As we can see in the
figure, all LSBD follow well an exponential profile characteristic
of disk-only galaxies, outside the area strongly affected by seeing
(r$>$4\arcsec). However, a few of the galaxies show a prominent bulge,
e.g. 3330-10 and 1116-11. In the central field - CTIO - the large pixel
size ($\sim$ 2\arcsec) could be masking the presence of a bulge, if
it exists.  However, three galaxies in the CTIO frame hint and evidence
for a bulge (4536-01, 5622-01 and 7731-01).

\subsection{Color profiles}

In order to analyze the color profiles of the 65 LSBD galaxies we have
obtained their magnitudes within concentric circles of growing radii
(starting from 3.5\arcsec and 2\arcsec radius for CTIO and LCO40 frames
respectively) using SExtractor.  For each galaxy, the profile for
the I band was subtracted from that for the V band to form the color
profile. Figure \ref{fig11} shows the color profiles of the 65 LSBD for
which color information is available. Most of the galaxies show flat color
profiles, i.e. very small gradients with an average for all galaxies
of $d[V-I]/dR=-0.014\pm 0.052$.  However, in a few cases we measured a
small inclination in the profiles. Galaxies 874-01, 1843-01, 3747-01,
1064-03 and 2674-06 become redder with increasing radius. This effect
was also found by Patterson \& Thuan (1996) for some dwarf irregular
galaxies and by Bremnes et al. (1998, 1999) for low-surface brightness
galaxies in the M81 and M101 groups. Other galaxies like 873-01, 2374-01,
4136-01, 5012-01 and 7022-01 become bluer with increasing radius. These
results are discussed in section 4.

\subsection{Projected number density of the LSBD galaxies}

Figure \ref{fig12} shows the logarithm of the surface number density (in
units of h$^{2}$ kpc$^{-2}$) as a function of the distance to the group
center (the centroid of the two central galaxies) for all LSBD galaxies
detected in the region of Dorado (the five LSBD galaxies detected in
LCO40 at $\mu_{lim}=26.5$ V mag/arcsec$^{2}$ were not included in this
plot, but if they are included the result is unchanged). The data for
the central region (CTIO) was divided in bins of $\sim 75$ h$^{-1}$ kpc
(0.25 degree) and normalized by the corresponding area. For the LCO40
data each field was plotted as one point in Fig. \ref{fig12}, where the
number of galaxies was normalized by the total area used in each field.

From this figure we can see that there is a clear central concentration
of LSBD, inside a radius of $\sim 250$ h$^{-1}$ kpc ($\sim$ 0.8 deg). At
a radius $r>250$ h$^{-1}$ kpc the projected number density is similar
to the values found in our control fields ($\sim 7$ galaxies per square
degree) and by Dalcanton et al. (1997). At $\sim 400$ h$^{-1}$ kpc
($\sim 1.3$ deg) and $\sim 700$ h$^{-1}$ kpc ($\sim 2.1$ deg) from the
center we found a slight enhancement in the projected number density
of LSBD galaxies. These density enhancements could be associated to the
early-type galaxies NGC1536 (field D05) and NGC1574 (field D11).  Two of
the three galaxies in the D11 field that may be associated to NGC1574
galaxy are dwarf irregulars and one is a dwarf elliptical galaxy.
Figure \ref{fig13} shows the projected distribution of all bright
galaxies with known redshift, like Fig. \ref{fig1}, but including the
LSBD galaxies detected in our survey. From the figure we can see that
the majority of the galaxies lie in the central region and five of the
nine galaxies in the outskirts of the group seem to be associated to
bright early-type galaxies.

We fit an exponential profile to the data shown in Figure \ref{fig12}
(for the three points with $r\le250$ h$^{-1}$ kpc) in order to find
the scale size and the central (extrapolated) galaxy number density
of the LSBD galaxy. We find a scale size of 270 h$^{-1}$ kpc ($\sim$
0.9 deg).  Alternatively, if we fit a power law of the type $\Sigma
\propto r^{-\beta}$ within a radius of 250 h$^{-1}$ kpc, we find that
$\beta = -1.09\pm0.32$.

\section{Discussion and Summary}

\subsection{Colors and color profiles}

The colors and color profiles of the LSBD may give us information on the
stellar content and other physical parameters as a function of radius
(for example, metallicity). The color distribution of the LSBD candidates
of the Dorado group range from $-0.3 < V-I < 2.3$ with a peak at $V - I=
0.98$ (Figure \ref{fig9}). The color range and peak value are in a good
agreement with the values obtained for the LSBD in the Virgo and Fornax
clusters (Impey et al. 1988, Bothun et al. 1991) and in the Pegasus and
Cancer spiral-rich clusters (O'Neil et al. 1997). About half the LSBD
population of Dorado is composed of blue galaxies with $V-I<1$ and 12\%
are very blue with $V-I<0.5$. These values are comparable, in percentage,
to those found by O'Neil et al. (1997) in the Pegasus and Cancer nearby
groups. As pointed out by these authors, the blue LSBD are important
because of the restrictions imposed on their star formation histories
and also because they could be the local counterparts of the distant
faint blue galaxies.

The average flat color profiles found for the LSBD in Dorado shown in
section 3.3 suggest that there is no star formation or, if it exists,
it is very weak.  The very blue colors in some galaxies is restricted
to the central regions and they can be explained by the existence of a
sporadic burst with a constant, but very low, star formation rate. The
star formation decreases or disappears in the outer parts of the galaxies.

Rich clusters like Coma do not contain a large population of blue
LSBD objects most probably because the evolution of these galaxies is
accelerated and they could not survive a long time in such a hostile
environment. The color distribution of the Coma dwarf galaxies peaks
at $B-R = 1.4$ and at fainter magnitudes they become bluer, to a mean
$B-R$ of $1.15$ (Secker et al. 1997).  These two peaks are much redder
than the peak in the Dorado group ($V-I=0.98$, corresponds to a typical
$B-R\sim0.5$). This is expected, since the Dorado group has a very low
density environment of bright galaxies.

Blue LSBD have been identified and are well studied in a number
of nearby groups and in the field (e.g. McGaugh \& Bothun 1994,
de Block et al. 1995, Impey et al. 1996, Pildis et al. 1997, O'Neil
et al. 1997). However, we still lack good information on red LSBD
galaxies since most of the previous works were done with photographic
plates. Practically all of the LSBD observed in photographic surveys
are blue since photographic data are biased towards finding blue
objects. About 12\% of the LSBD sample in this work shows colors $V-I>1.5$
mag. For some galaxies the colors are very red (e.g. 779-01, 4536-01 and
3330-10 with $V-I>1.8$) with a flat color profile (see Fig. \ref{fig11}).
The red colors found for these galaxies could not be explained by
photometric errors (the errors are $<0.3$ mag, see section 2.6), by
interstellar absorption (the color correction is less than 0.02 mag)
or by internal reddening (there is no evidence for a high dust content
in the LSBD galaxies, O'Neil et al. 1997).

Detection of very red LSBD galaxies in nearby poor groups is not new.
Recently, in a survey of low-surface brightness galaxies in the
Pegasus and Cancer clusters taken by O'Neil et al. (1997), a significant
population of red LSBD galaxies was discovered. About 20\% of the galaxies
in their sample have colors $V-I>1.5$. The percentage of red galaxies
in Dorado is similar ($\sim$ 12\%). One probable scenario to explain the
colors of these galaxies, discussed by O'Neil et al. (1997), is that these
objects ``underwent starburst early in their existence, consuming most
of the galaxies gas and leaving them with an old stellar population''.

Some examples of very red LSBD ($V-I>1.8$) in the Dorado group are
779-01, 4536-01 and 3330-10. Due to the poor resolution of the CTIO
image, it was impossible for us to obtain a morphological classification
for 779-01 and 4536-01.  However, the 3330-10 galaxy was classified
as a dIrr galaxy. This galaxy is a special case because it is very
red ($V-I\sim2.2$) and shows an irregular morphology, with a large
disc and strong bulge.  A number of very red low surface brightness
galaxies detected by 0'Neil et al. (1997) were recently surveyed in HI
by O'Neil et al. (2000). They showed that a few of them are gas-rich
low-surface brightness galaxies with no massive star formation and with
$M_{HI}/L_{B}>9$. We have no information about the HI content of the
3330-10 galaxy and we can only speculate that this galaxy could have
properties similar to those of the galaxies discovered by O'Neil et
al. (2000) in their survey. Additional HI observations are needed to
confirm this point.

\subsection{Projected number density of the LSBD galaxies}

The clustering properties of the dwarf galaxy population is an important
tool for cosmology. The space distribution of the dwarf satellites
around bright galaxies could indicate how luminous and dark material
are distributed at different scales. Moreover it may be a tracer of the
dynamical evolution of the dark halos around bright galaxies.

As determined in Section 3.4, the central galaxy concentration dies off
outside a distance to the group center of $\sim$ 250 h$^{-1}$ kpc. The
radial distribution (Fig. \ref{fig12}) is in general agreement with the
results obtained by Mulchaey \& Zabludoff (1999) for five X-ray groups,
lthough they cannot be compared in detail given the different magnitudes
considered in this and their studies (we consider galaxies with $-16<
M_V <-11$ and they considered galaxies with $-17 < M_V <-15$ for H$_0$
$=$ 75 km s$^{-1}$ Mpc $^{-2}$ and $V-R\sim0.6$).

It is well known that dwarf elliptical galaxies show a stronger clustering
around early-type galaxies (e.g. Vader \& Sandage 1991) than around
late-type galaxies (e.g. Lorrimer et al. 1994) and are more common in
large and rich environments. The Dorado group has two bright early-type
galaxies (one E1 and one S0) in the center, separated by only 58 h$^{-1}$
kpc, and has no other prominent early-type object in its outskirts. In
a radius of $\sim0.5$ h$^{-1}$ Mpc from the central galaxies, only four
galaxies with absolute magnitudes brighter than M$_V = -20$ are present,
and they are all late-type objects. In order to analyze if the LSBD
galaxies detected in Dorado group are associated to the bright central
early type NGC1549 and NGC1553 galaxies of Dorado or to the potential
well of the group, we plotted the distribution of all LSBD detected in our
survey around three different positions: {\it 1)} the E1 galaxy NGC1549,
{\it 2)} the centroid of the two central group galaxies and {\it 3)}
the S0 galaxy NGC1553. We found no significant difference between these
three plots. The central concentration detected suggests the galaxies
are true dwarf galaxies of the Dorado group but we cannot decide from
the exercise above if they belong to the group or to the two central
early-type galaxies.

To further test this point, we compared our result with those obtained for
satellites of isolated bright elliptical galaxies.  The study of Vader \&
Sandage (1991) found that the density of satellites, if fit to a function
$\Sigma \propto r^{-\beta}$, with $\beta = -1.22$. This agrees with our
value of $-1.09$.  Furthermore, the distribution of satellites around
the center of Dorado is similar in extent to that found around the the
X-ray bright elliptical galaxy NGC 1132 studied by Mulchaey and Zabludoff
(1999).  These comparisons would seem to suggest that the LSBD galaxies
in the Dorado group are most probably physically associated with the
two central galaxies. However, since these galaxies strongly dominate
the potential well of the group, this may be equivalent to saying that
the satellites we measured are associated with the overall potential
of the group.  An alternative interpretation of these results would be
that isolated elliptical galaxies and groups have similar distributions
of satellites around them perhaps because all bright field elliptical
galaxies may be merged groups (as may be the case for NGC 1132 discussed
by Mulchaey and Zabludoff 1999).

\subsection {LSBD, tidal interacting feature or Galactic cloud?}

We discovered an extended low-surface brightness ``cloud'' $\sim$
10\arcmin (about 50 h$^{-1}$ kpc) to the NW of the central NGC1549 galaxy
(galaxy L02 in Table \ref{tab3}, see Section 2.7.2).  The object, shown
in Fig. \ref{fig14} , has a smooth appearance, like a dwarf elliptical
galaxy. However, it is much larger than all the other satellite candidates
($\sim$ 10 $\times$ 5 h$^{-1}$ kpc) and it is very close to one of the
central galaxies.

Bremnes et al. (1998) has detected several low-surface brightness clouds
in the direction of M81 which overlap with detections in the IRAS 100
$\micron$ map, suggesting that these may indeed be cirrus structures in
our own Galaxy.  However, in the case of Dorado, the IRAS 100 $\micron$
map of the region shows no similar features. Moreover, the ``cloud''
in Dorado is much closer to a bright galaxy than in the case for the
clouds in the M81 group.

An alternative explanation would be that this object is the residual
of a past galaxy-galaxy interaction or a disrupted dwarf galaxy like
Sagittarius in the Local group. Higher resolution imaging of this object
may reveal its nature.

\subsection {Summary}

Summarizing, we have detected 69 low-surface brightness dwarf galaxies
as candidate dwarf members of the nearby group of galaxies Dorado. Our
CCD survey covered $\sim 2.6$ square degrees of the group. The galaxies
were chosen based on  their sizes and magnitudes at the average limiting
isophote of $26$ V mag/arcsec$^{2}$. We showed, using Monte Carlo
simulations, that we are able to detect possible dwarf members in Dorado
with $-10.5>M_{V}>-17$ with $22.5<\mu_{0}<25.5$ V mag/arcsec$^{2}$ with
a completion fraction $f>80$\%. Four new large low-surface brightness
galaxies previously undetected by standard techniques were detected
using a smoothing technic. One of these galaxies is a large low-surface
brightness ``cloud'' with an irregular shape much more extended than the
other dwarf galaxies in the survey.  The color distribution for the 65
LSBD galaxies with color information varies from very blue to very red
with a peak at $V-I = 0.98$.  The color profiles of the galaxies show
very small gradients (on the order of the photometric errors). In some
cases we measured a small inclination in the profiles. The reddest galaxy
(3330-10) shows an irregular morphology, a strong bulge, a large halo
and a flat color profile. This galaxy could be a HI-rich galaxy like
others of the same type found in the Pegasus and Cancer clusters by
O'Neil et al. (2000). The projected surface density shows a significant
excess of LSBD galaxies within $\sim 250$ h$^{-1}$ kpc ($\sim$ 0.8 deg)
from the center of the group. Comparisons with others results obtained
from X-ray groups and for the isolated elliptical galaxy NGC1132 suggest
that the Dorado LSBD are probably physically associated with the overall
potential well of the group.

\acknowledgements The authors are grateful to the Directors of CTIO and
Las Campanas Observatory for generous allocation of telescope time. ERC
acknowledges support for this work provided by FAPESP PhD fellowship Nr.
96/04246-7. CMdO is grateful to E. Bertin for several discussions about
the use of the SExtractor package. L.I. was supported partially by
Proyecto FONDECYT Nr. 8970009 and a Guggenheim Foundation award. This
work benefitted from the use of the NASA/IPAC Extragalactic Database
(NED), which is operated by the Jet Propulsion Laboratory, California
Institute of Technology, under contract with the National Aeronautics
and Space Administration.

\newpage

%
%

\figcaption[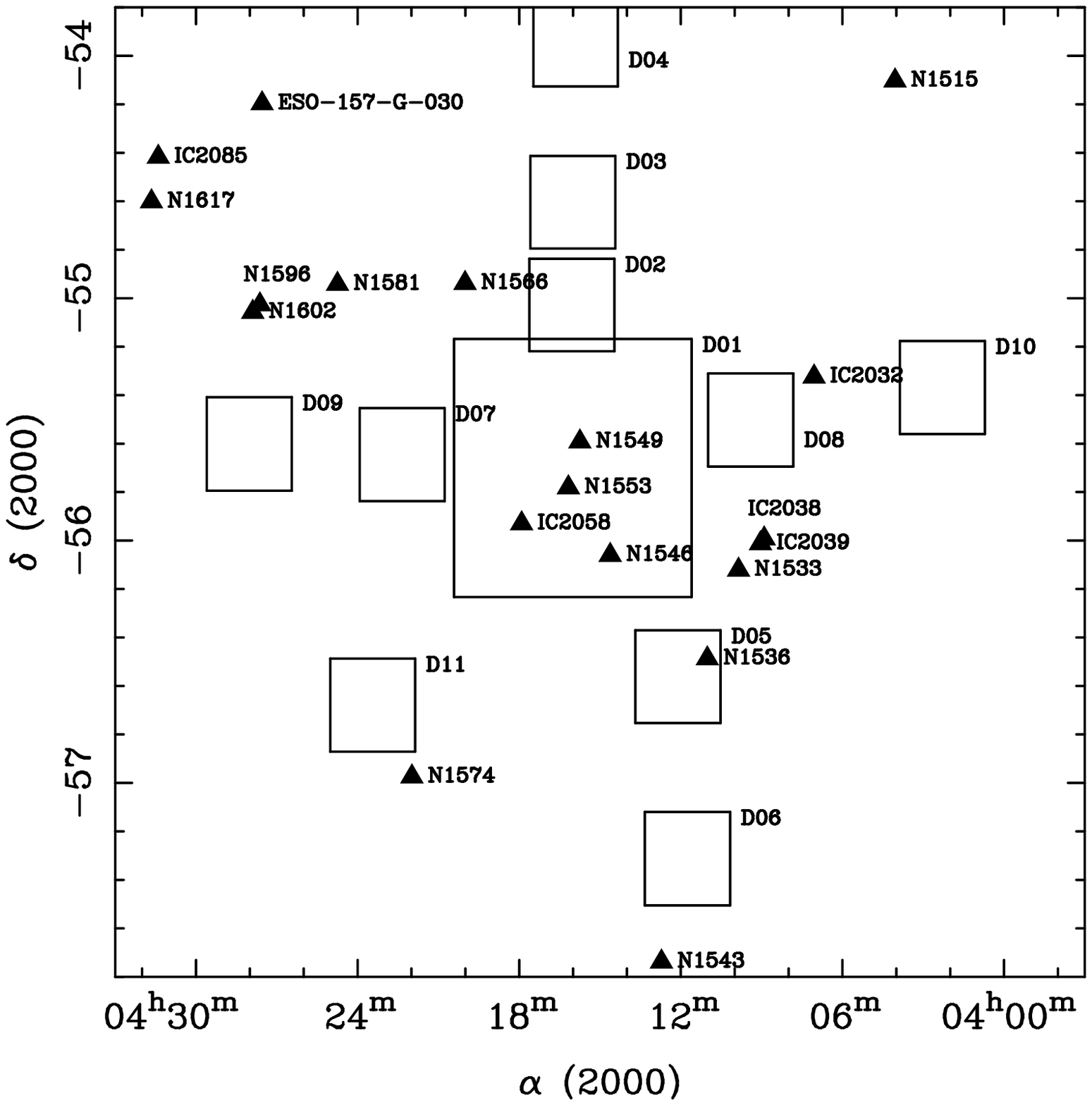]{Locations and sizes of the observed CCD
frames in the region of the Dorado group. The small squares indicate the
LCO40 frames (24\arcmin $\times$  24\arcmin) and the large square the
CTIO frame (1.15\arcdeg $\times$ 1.15\arcdeg). The triangles indicate all
galaxies with known velocities $cz<2000$ km/s, taken from the NED. North
is up and East is to the left. \label{fig1}}

\figcaption[carrasco.fig2.eps]{A V image of the central region of the
Dorado group ($\approx 1^{\circ} \times 1^{\circ}$) taken with the 0.9
m Schmidt telescope at  CTIO, Chile. The brightest galaxies in the frame
are identified by their NGC or IC numbers. At the bottom of the figure,
a mosaic of selected low-surface brightness dwarf galaxies  detected in
this region is shown. The small squares in the image indicate the location
of the galaxies in the area. The photometric and profile fit parameters
of these galaxies  are listed in Table 3 (see section 2.5). \label{fig2}}

\figcaption[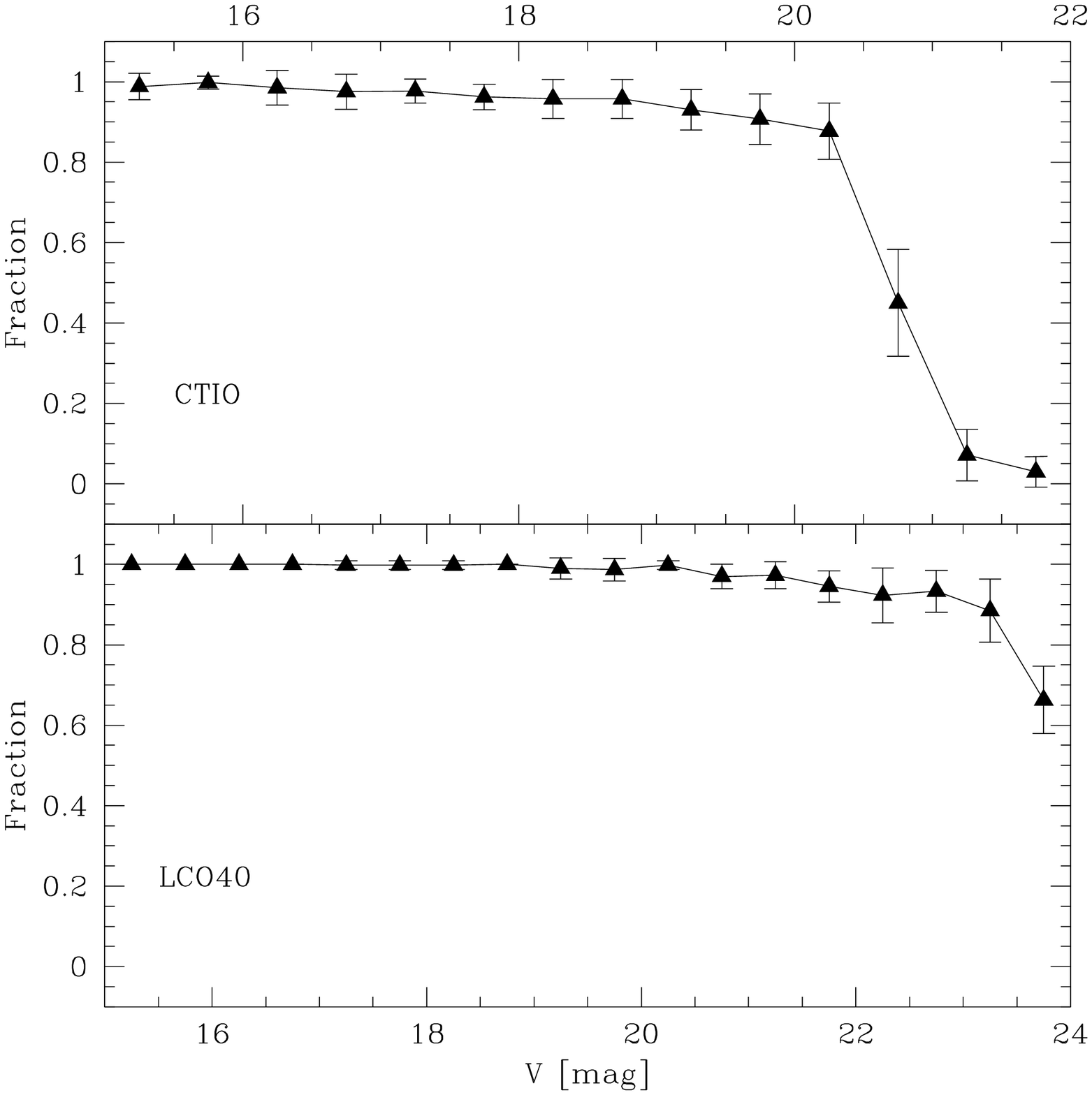]{Completeness fraction versus the input
total magnitudes for the simulated point sources.  The completeness
limit for point sources with ``stellarity index'' $\ge 0.9$ is $\sim
20.5$ for the CTIO frame (upper panel) and $\sim 23$ for LCO40 frames
(lower panel).\label{fig3}}

\figcaption[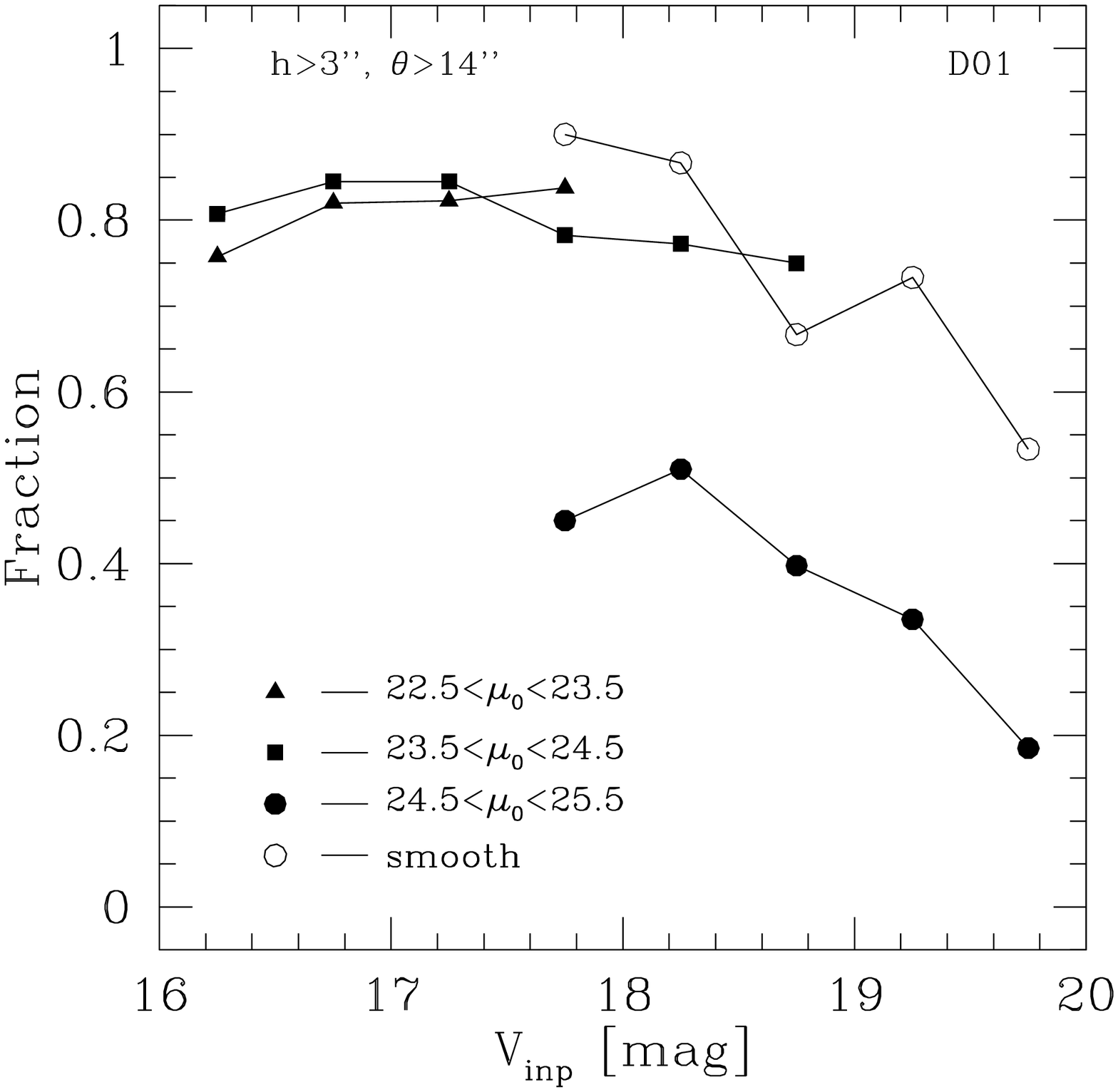,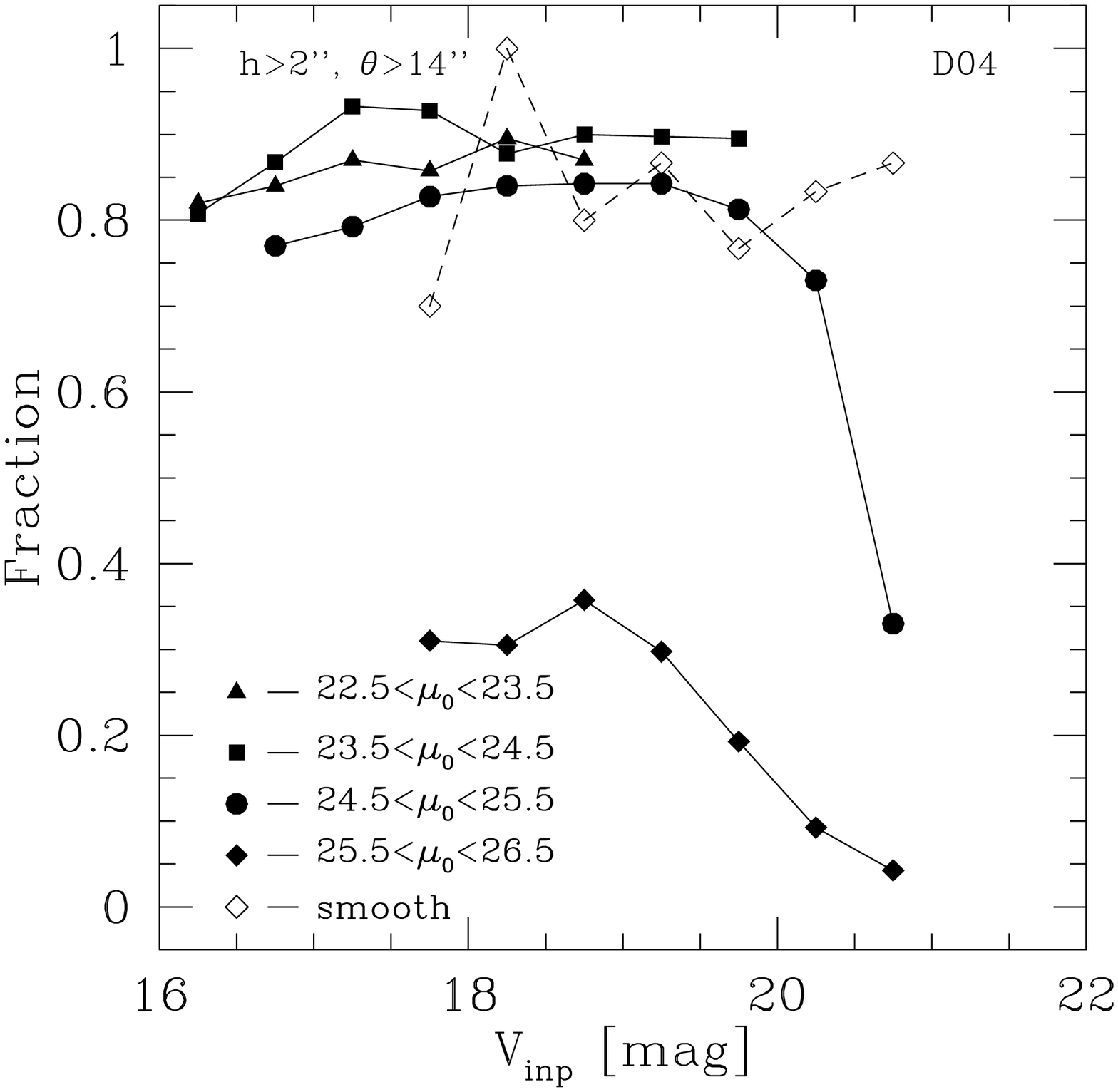]{Completeness
fraction versus the input total magnitudes for LSBD simulated galaxies
added to the observed frames. The upper and lower panels show the
results for the simulations done using the CTIO and LCO40 frames
respectively. \label{fig4}}

\figcaption[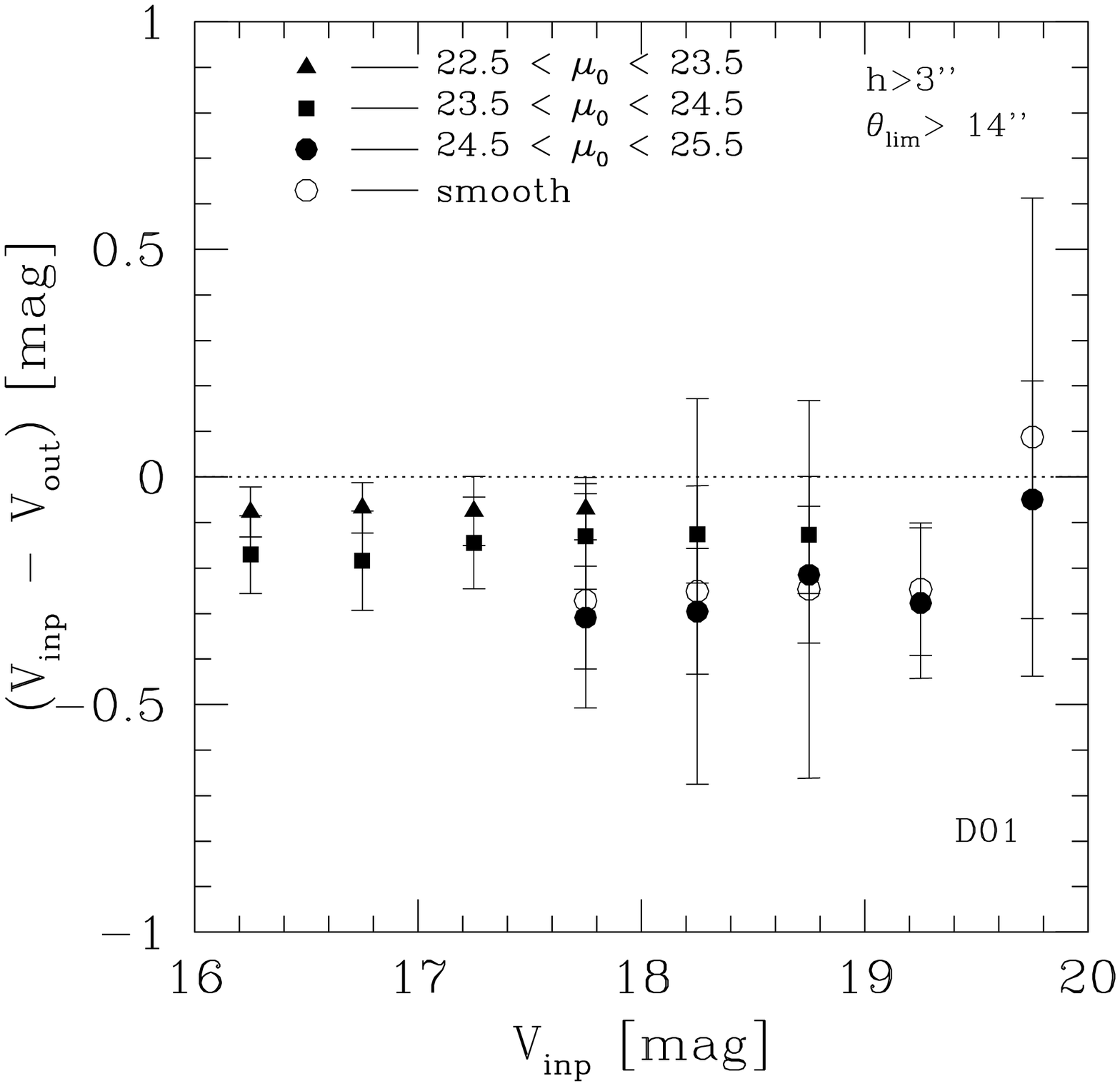,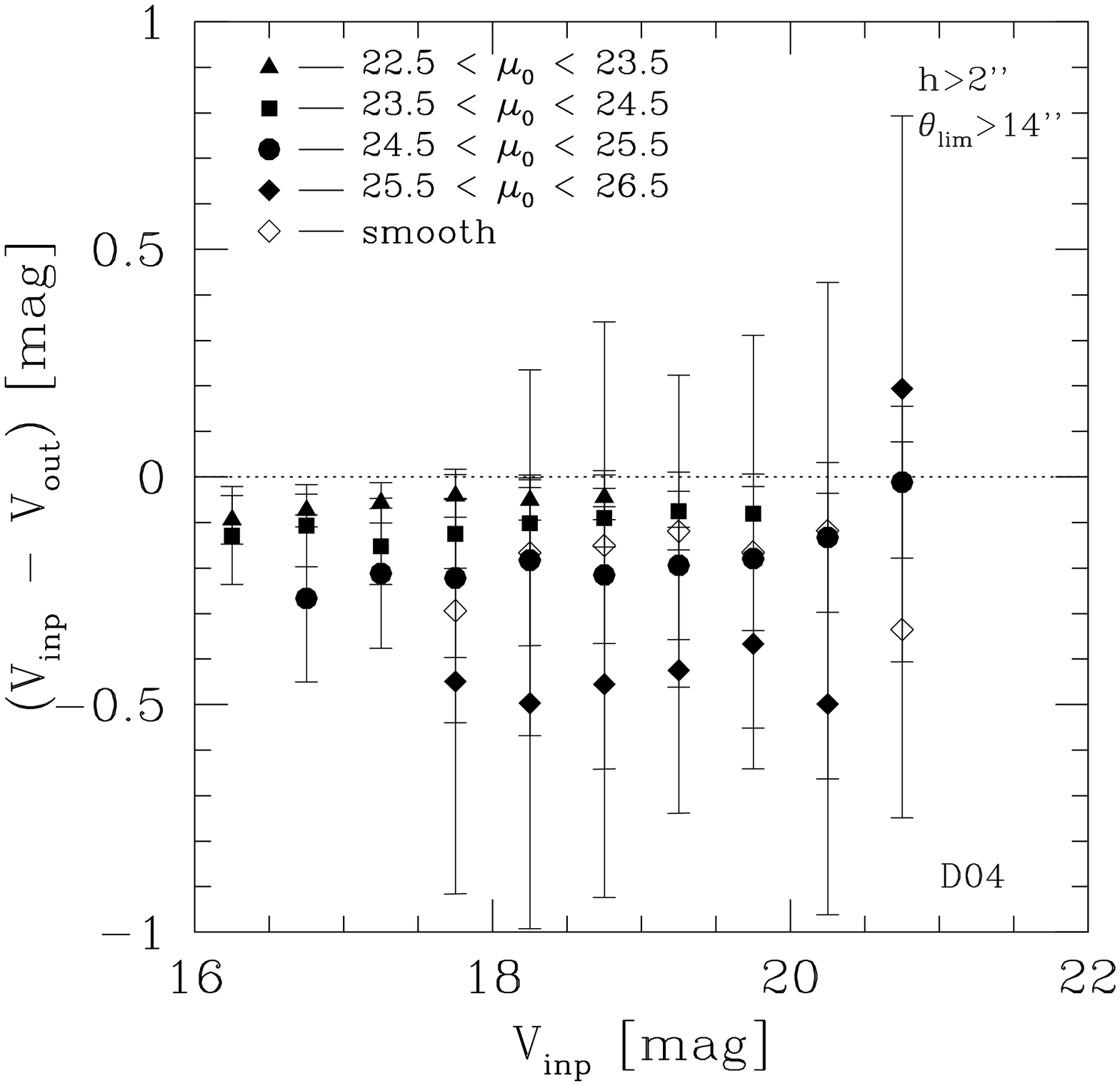]{Differences between
the input and output magnitudes obtained from the LSBD  Monte Carlo
simulations for the fields D01 (upper panel) and D04 (lower panel). In
the upper panel (CTIO), $\sigma(\Delta m) < 0.3$ mag ($\Delta m =
m_{inp} - m_{out}$) for LSBD with $22.5<\mu_{0}<24.5$  and $16<V_{t}<21$
(completeness $f>80$).  In the lower panel (LCO40), $\sigma(\Delta m) <
0.3$ for LSBD with $24.5<\mu_{0}<25.5$ and $18<V_{t}<20$ (completeness
$f>80$). No real galaxies were found in the last bins of surface
brightness for $V<18$. We, therefor, do not plot the corresponding
simulations in this figure to avoid confusion.  \label{fig5}}

\figcaption[carrasco.fig6.eps]{Extended LSBD detected using the
smoothing technique. The upper left panel shows the masked image of a
galaxy that could not be found by SExtractor at a first pass (section
2.3). The upper right panel shows the smoothed image of the same
galaxy after filtering. This galaxy has a magnitude of 18.5, a scale
length h=13.2\arcsec and a central surface brightness $\mu_{0}=24.8$
V mag/arcsec$^{2}$ (see galaxy L01 in Table \ref{tab3}). \label{fig6}}

\figcaption[carrasco.fig7.eps]{Simulated galaxy with $V\sim18.5$ and
$25.5<\mu_{0}<26.5$ in one of the  LCO40 fields (upper left panel). All
high-surface brightness objects with a threshold  over 3$\sigma$ are
masked (upper right panel). The smoothing filter is applied in order
to enhance the galaxies that are at the limit of detection (lower left
panel). After smoothing, SExtractor is run in order to detect the extended
LSBD (lower right panel). At the limit of detectability of our images,
the completeness increases to $\sim 80$\%, using the technique (open
circles in figures \ref{fig4} and \ref{fig5}) . \label{fig7}}

\figcaption[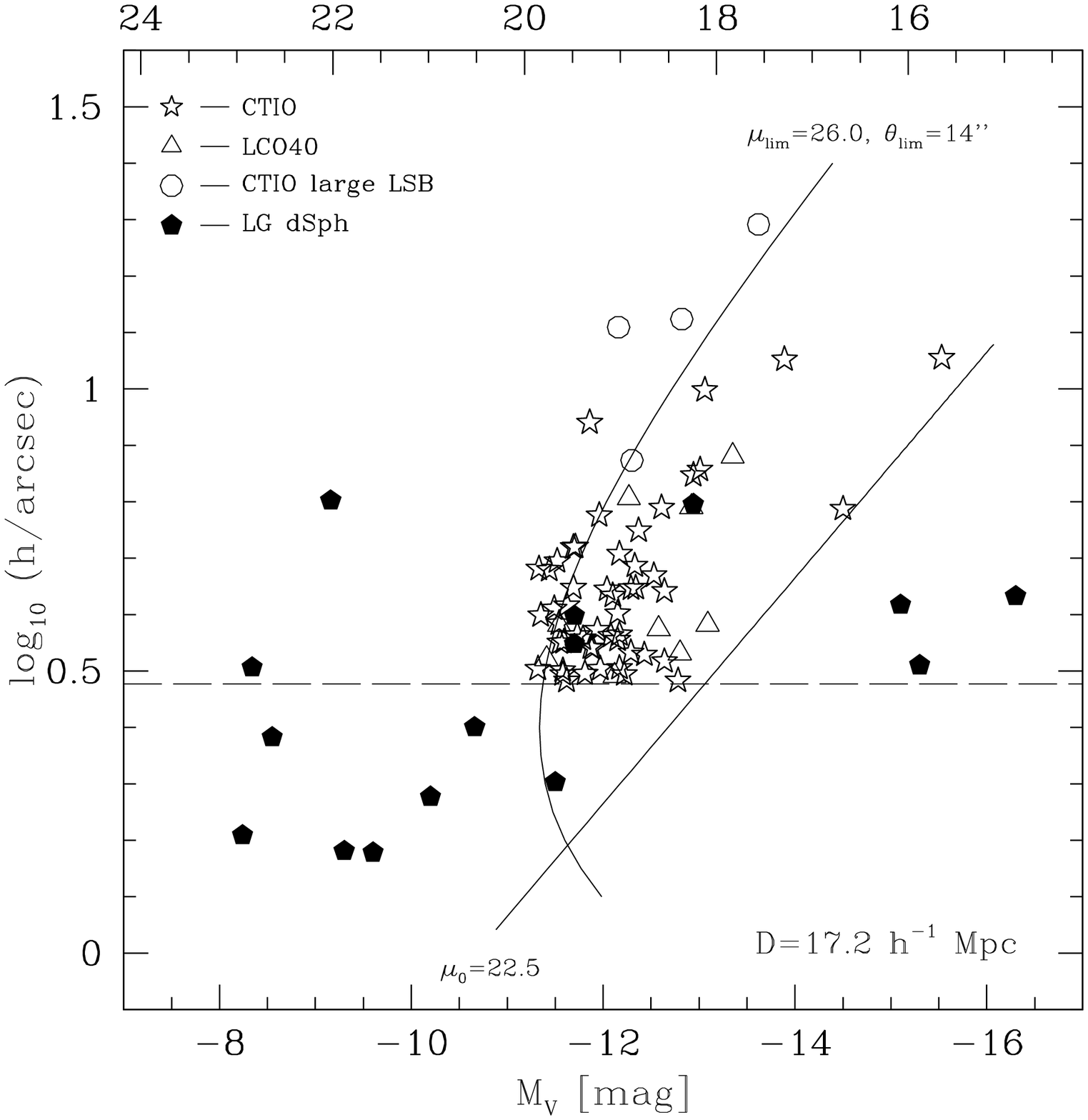,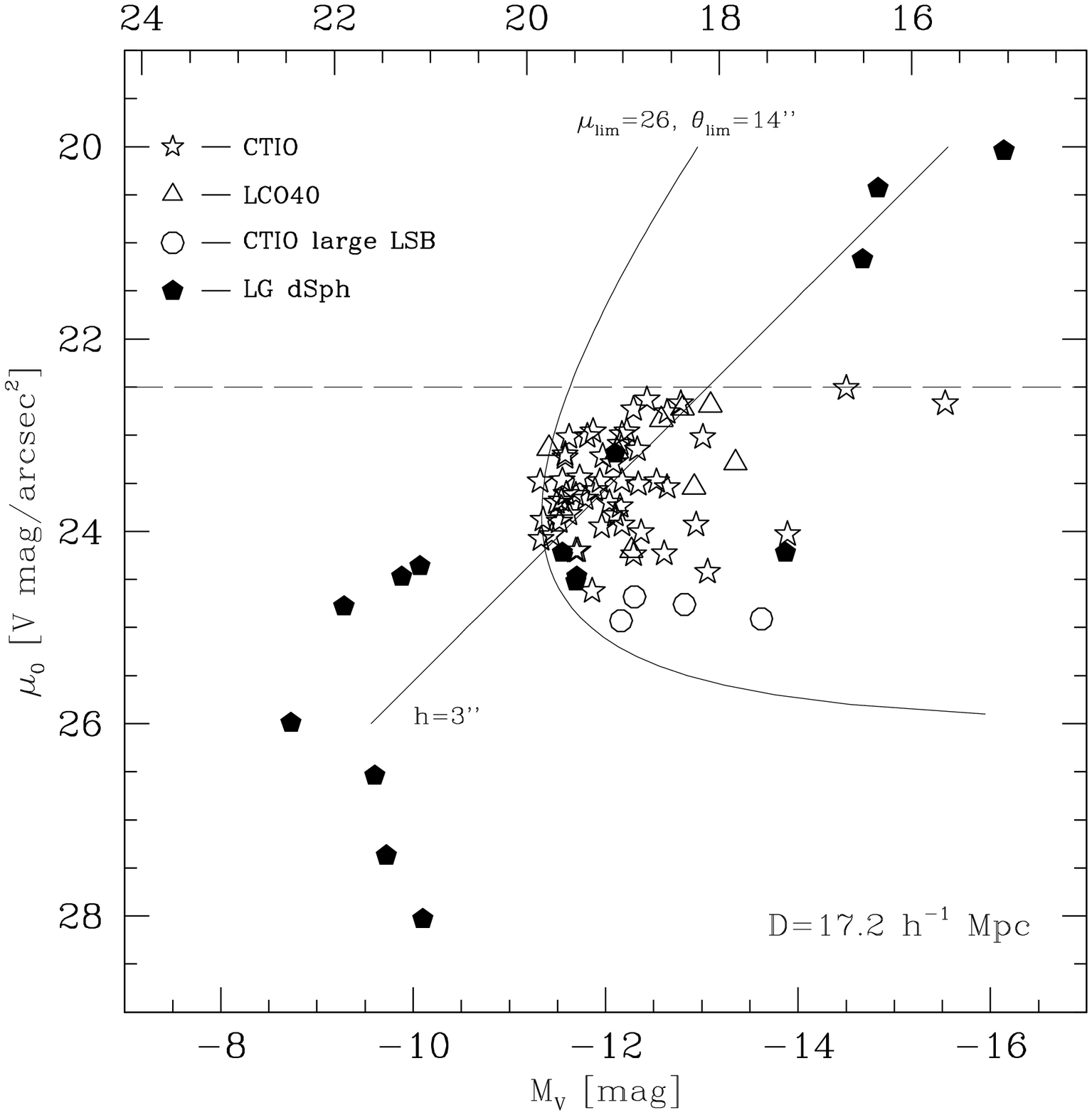]{Extrapolated central
surface brightness (lower panel) and scale length (upper panel) versus
total magnitude for LSBD galaxies detected in the Dorado group region
with $\mu_{0}>22.5$ V mag/arcsec$^{2}$, $h\ge2$'' (3'' for the CTIO frame)
and diameter $\ge14$''. The open stars and triangles are the LSBD galaxies
detected in the CTIO and LCO40 frames respectively. The open circles are
the large low-surface brightness galaxies detected using the smoothing
method of section 2.7. For comparison, we mark in the plot the Local
Group dSphs galaxies (filled symbols) redshifted to the Dorado distance
(17.2 $h^{-1}$ Mpc). The dashed lines are the central surface brightness
and scale length cuts used to select the LSBD (22.5 mag/arcsec$^{2}$
and 3\arcsec respectively). The solid lines represent the limiting
diameter given by equation (4) with $\mu_{lim}=26$ mag/arcsec$^{2}$
and by the relation between the total magnitude, central surface
brightness and the scale length for galaxies with exponential profiles
($m_{t}=\mu_{0}-5\log(h)-2.5\log(2\pi)$) \label{fig8}}

\figcaption[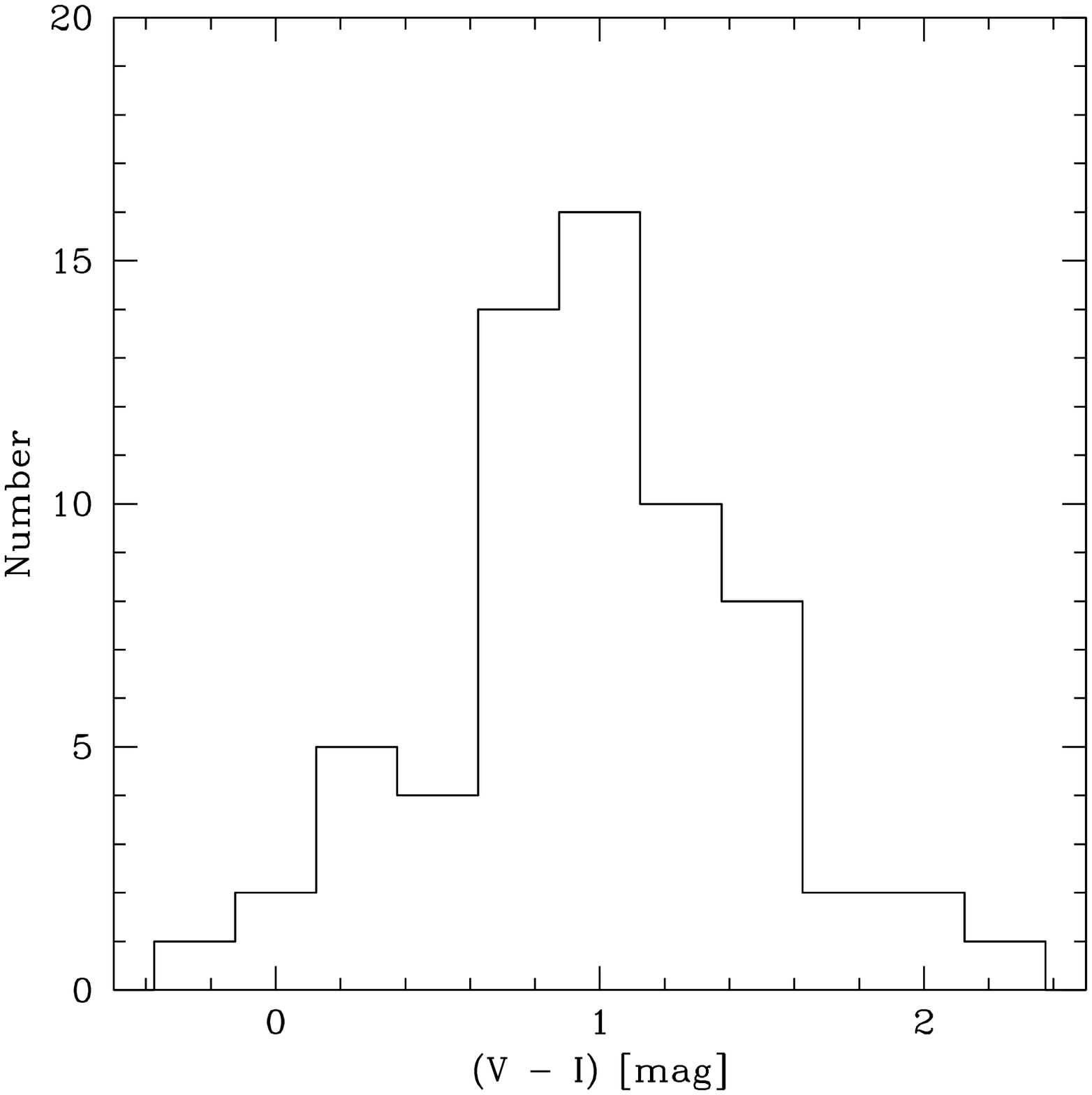]{Histogram of the color distribution of
the 65 LSBD galaxies with a color information. The LSBD galaxies in
Dorado show colors between $-0.3<V-I<2.3$ with a peak at $V-I=0.98$.
\label{fig9}}

\figcaption[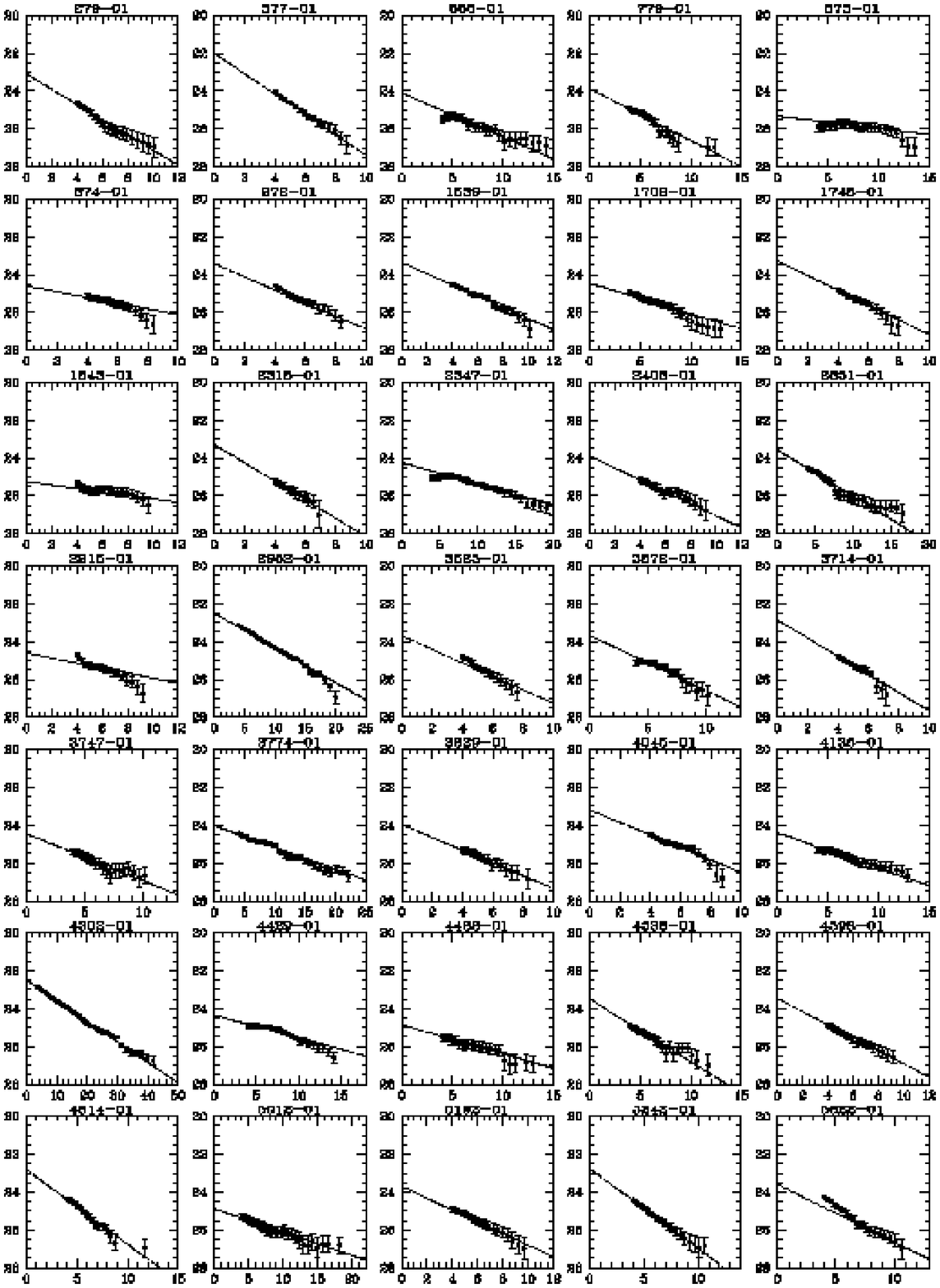,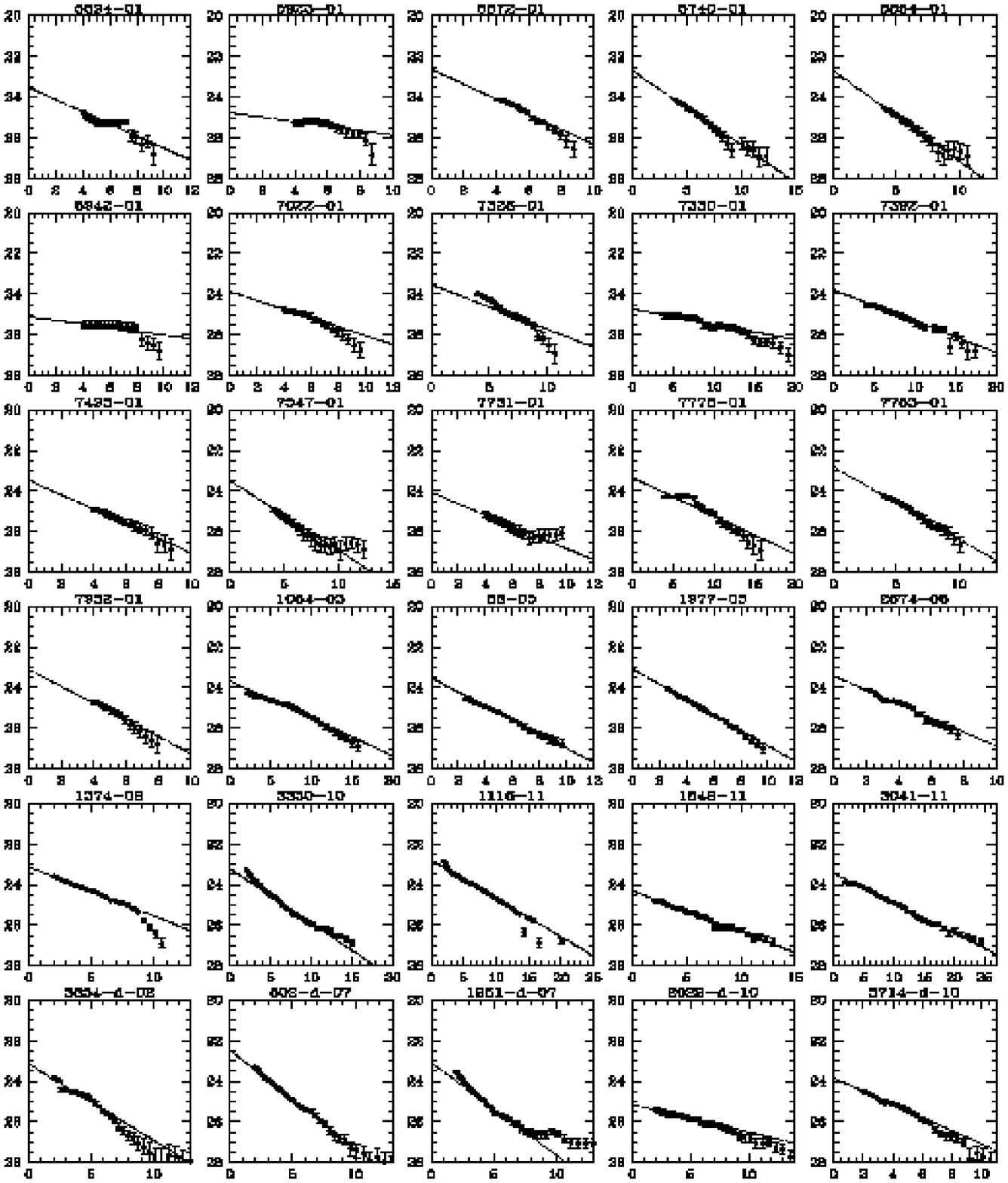]{Surface brightness
profiles of the 65 LSBD galaxies detected in the region of the Dorado
group (V filter). The dashed lines represent the exponential fits as
described in the text. The radii are the semi-major axis of the elliptical
apertures of the galaxies. \label{fig10}}

\figcaption[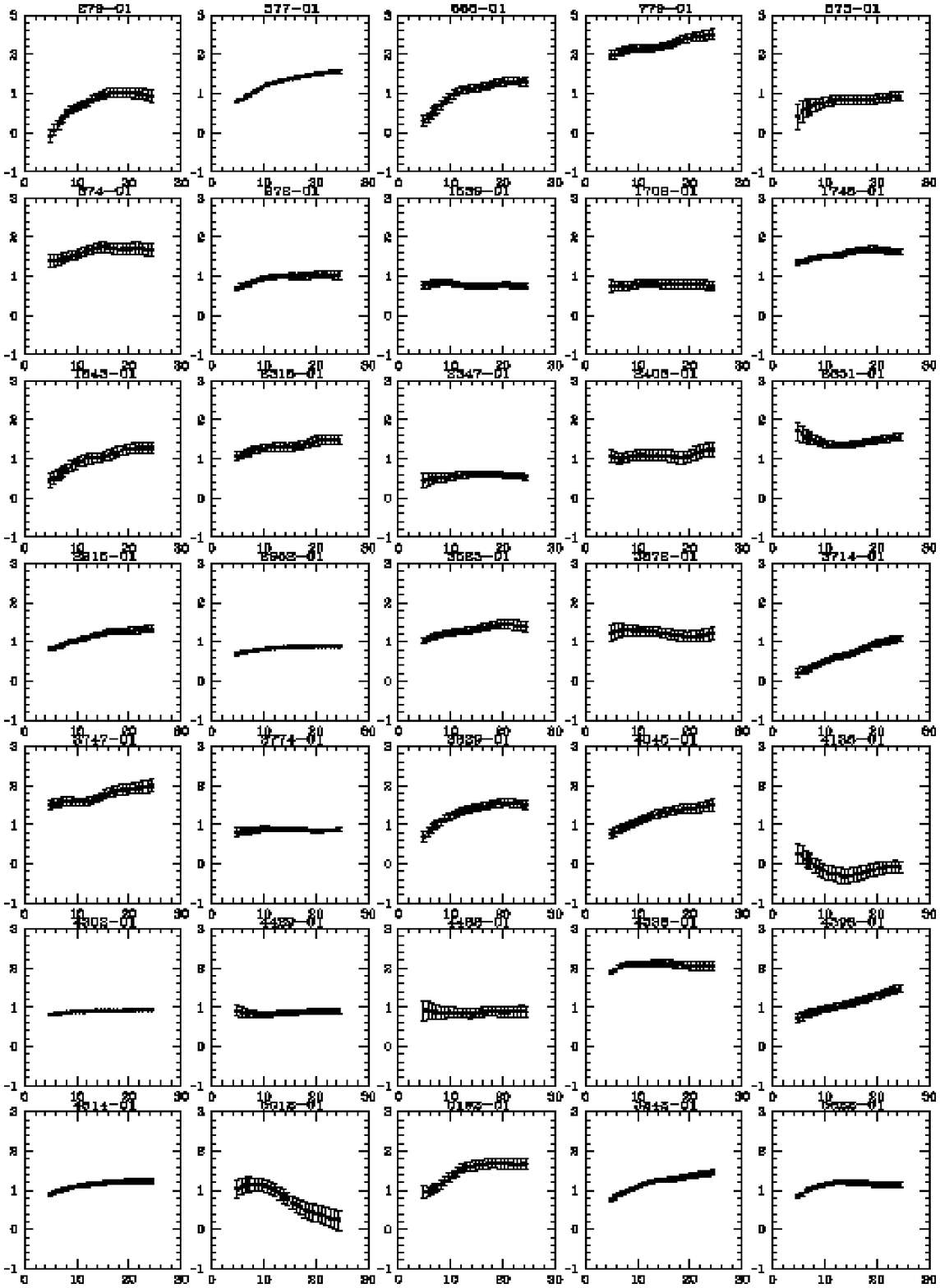,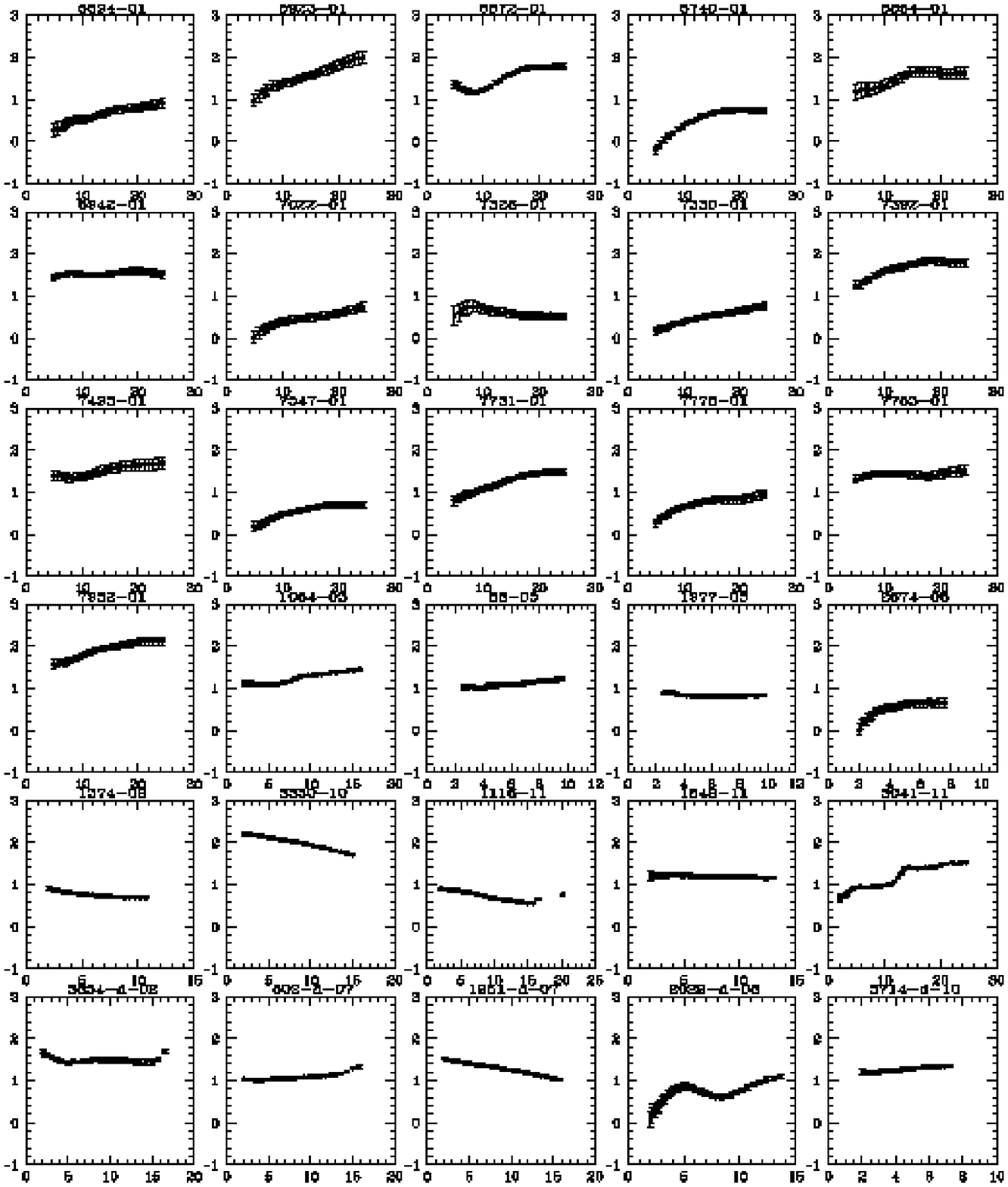]{Radial color
profiles $V-I$ for the 65 LSBD.  The color profiles were constructed
subtracting the profile for the I band from that for the V band. The
majority of the galaxies show a flat color profile with a gradient
$d[V-I]/dR=-0.014\pm0.052$. \label{fig11}}

\figcaption[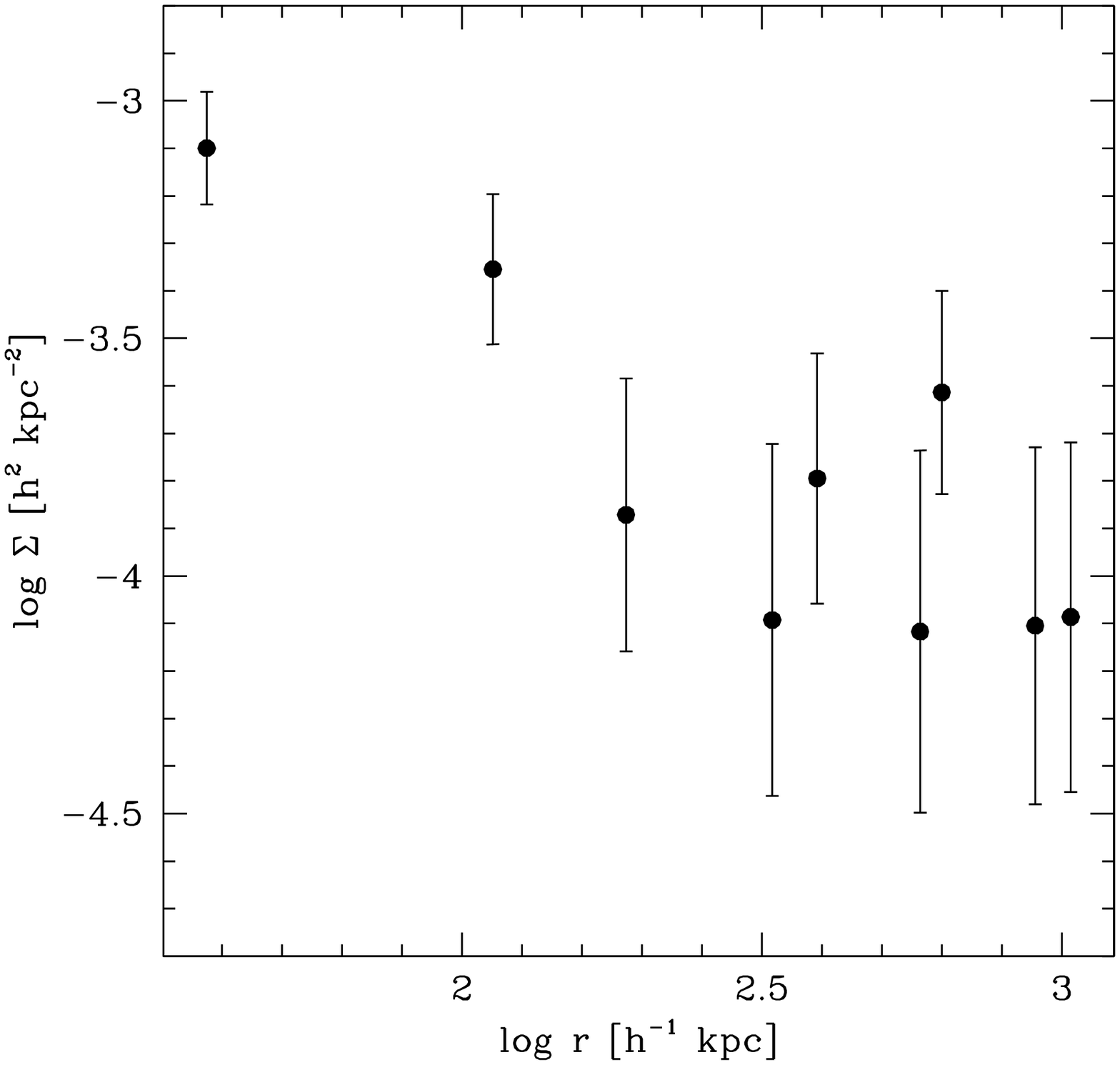]{Surface density as a function of the
radius (in logarithmic scales) for all LSBD galaxies detected in the
region of the Dorado group. The data for the central region was divided
in bins of $\sim$ 75 h$^{-1}$ kpc (0.25 degree). The radial distance
was measured from the centroid of the two central group galaxies NGC1549
and NGC1553. \label{fig12}}

\figcaption[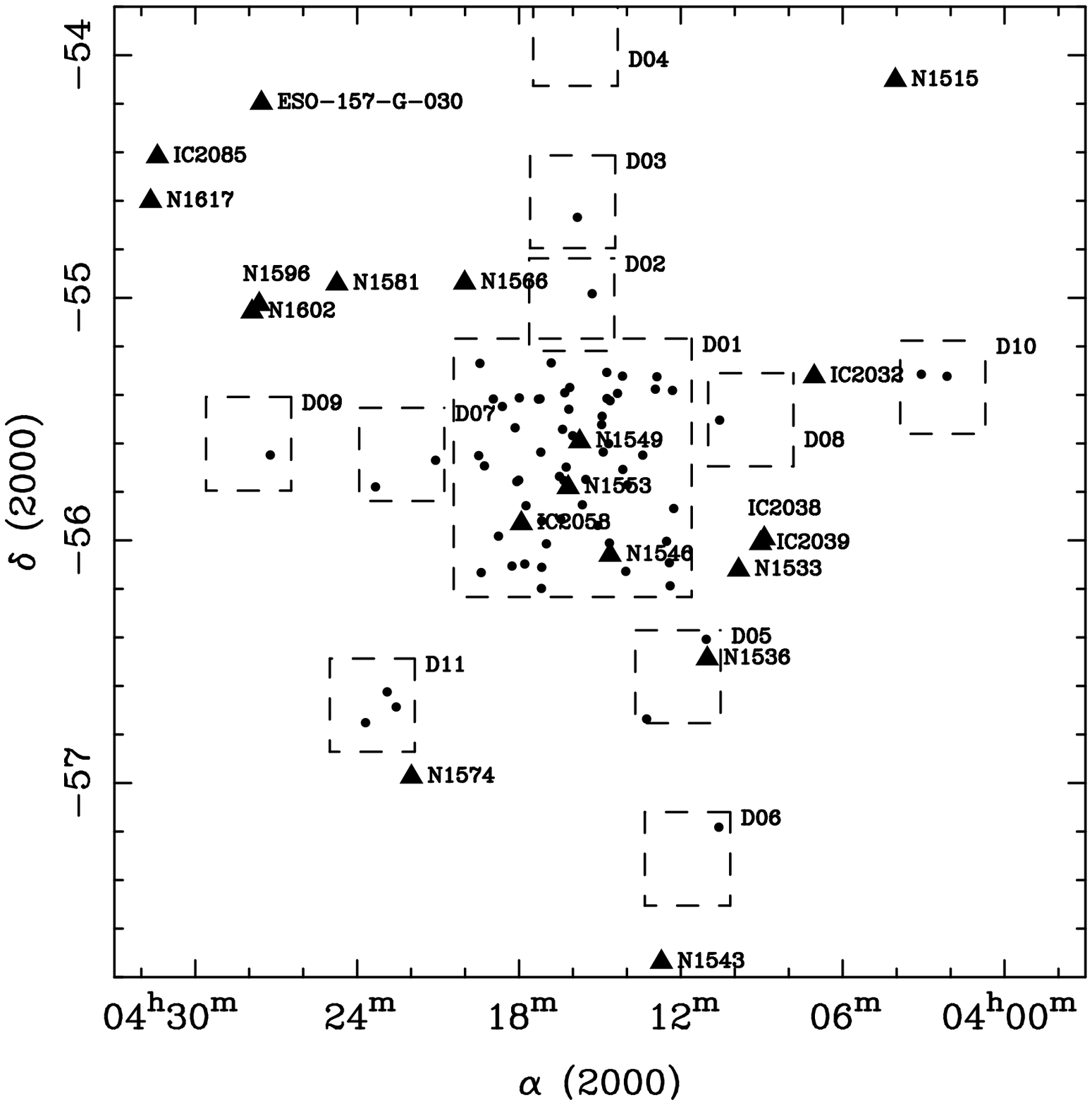]{Projected distribution of all bright
galaxies with known redshift, like Fig. \ref{fig1}, but including the
LSBD galaxies detected in our survey (small circles). \label{fig13}}

\figcaption[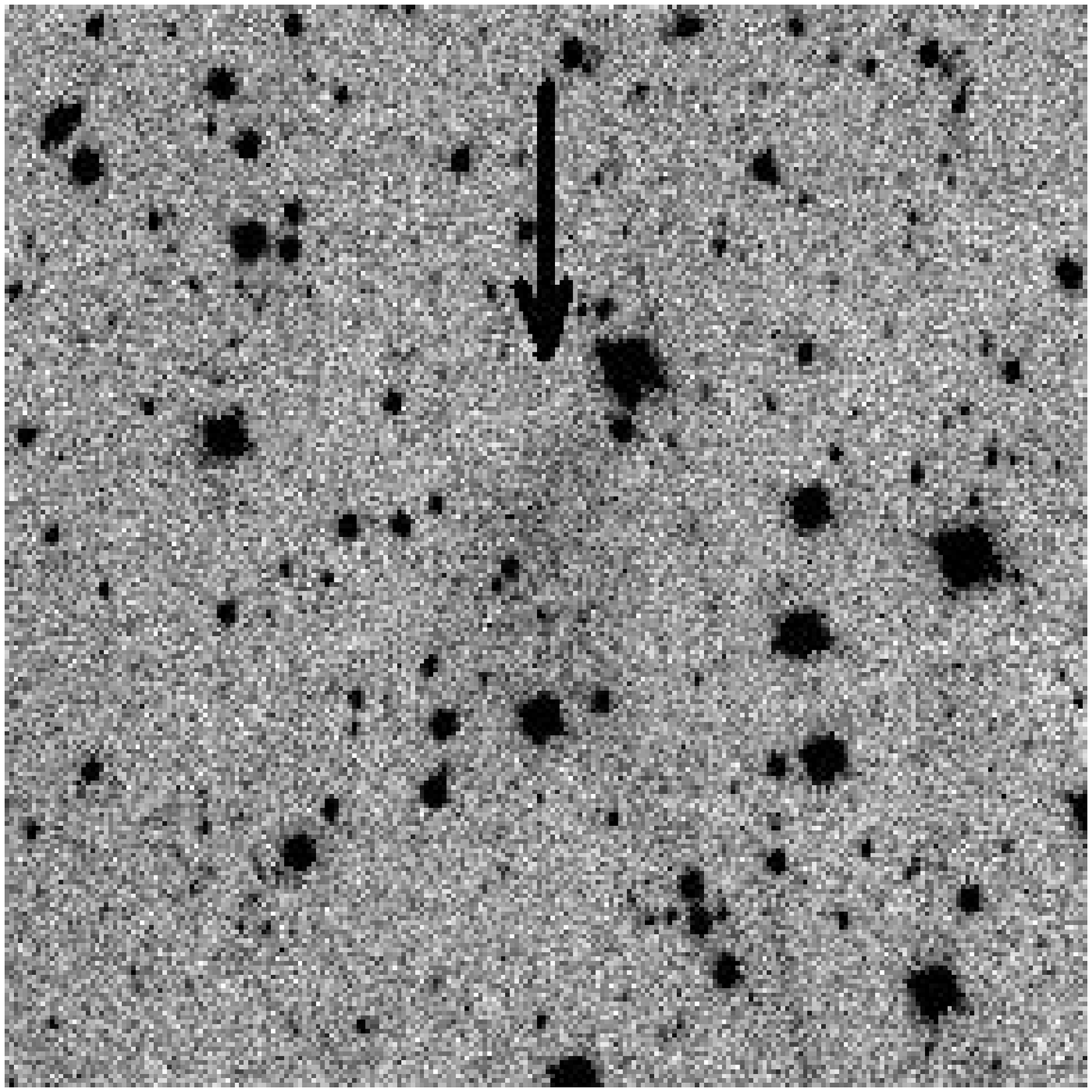]{CCD Image of the extended low-surface
brightness ``cloud'' located $\sim$ 10\arcmin NW from the central galaxy
NGC1549. The size of the ``cloud'' is $\sim$ 2\arcmin $\times$ 1\arcmin
(10 $\times$ 5 h$^{-1}$ kpc). The size of the figure is 8\arcmin $\times$
8\arcmin. North is up, East is to the left. \label{fig14}}
\newpage

\begin{deluxetable}{ccccrrr}
\tablenum{1}
\tablecolumns{7}
\tablewidth{0pc}
\tablecaption{Observation log \label{tab1}}
\tablehead{
\colhead{Field Id}&
\colhead{$\alpha(2000)$}&
\colhead{$\delta(2000)$}&
\colhead{Filters}&
\colhead{Exposure}&
\colhead{A$_{tot}$}&
\colhead{A$_{used}$}\\
\colhead{}&
\colhead{$^{h}$ $^{m}$ $^{s}$}&
\colhead{$^{\circ}$ $^{'}$ $^{''}$}&
\colhead{}&
\colhead{[sec]}&
\colhead{[arcmin$^{2}$]}&
\colhead{[arcmin$^{2}$]} \\
\colhead{(1)} &
\colhead{(2)} &
\colhead{(3)} &
\colhead{(4)} &
\colhead{(5)} &
\colhead{(6)} &
\colhead{(7)}}
\startdata
D01 & 4 16 00.5 & -55 42 03 & V,I & $6 \times 600$  & 4163.86 &3634.79 \\
D02 & 4 16 02.8 & -55 01 41 & V,I & $3 \times 1200$ &  534.42 & 481.79 \\
D03 & 4 16 00.5 & -54 36 15 & V,I &                 &  535.52 & 488.69 \\
D04 & 4 15 54.4 & -53 56 06 & V,I &                 &  532.11 & 482.58 \\
D05 & 4 12 06.3 & -56 33 44 & V,I &                 &  537.11 & 492.29 \\
D06 & 4 11 45.1 & -57 18 45 & V,I &                 &  540.61 & 516.76 \\
D07 & 4 22 20.6 & -55 38 43 & V,I &                 &  537.93 & 499.00 \\
D08 & 4 09 24.4 & -55 30 09 & V,I &                 &  539.00 & 492.89 \\
D09 & 4 28 01.3 & -55 36 06 & V,I &                 &  539.84 & 502.28 \\
D10 & 4 02 17.4 & -55 22 07 & V,I &                 &  538.46 & 481.53 \\
D11 & 4 23 26.4 & -56 40 44 & V,I & $3 \times 900$  &  536.88 & 486.64 \\
B3  & 3 38 25.3 & -25 15 05 & V,I & $3 \times 1200$ &  517.42 & 451.65 \\
B4  & 3 38 22.5 & -49 59 24 & V,I &                 &  536.16 & 477.55 \\
\tablecomments{D01: CTIO observation; D02 to D11 and B3, B4: LCO40 observations}
\enddata
\end{deluxetable}

\begin{deluxetable}{lcll}
\tablenum{2}
\tablecolumns{4}
\tablewidth{0pc}
\tablecaption{Transformation Equations\label{tab2}}
\tablehead{
\colhead{Run} &
\colhead{Night} &
\colhead{Transformation equations\tablenotemark{(a)}} &
\colhead{rms} \\
\colhead{(1)} &
\colhead{(2)} &
\colhead{(3)} &
\colhead{(4)}}
\startdata
CTIO & Nov-1-96 & $V = v - 4.904\pm0.019 - 0.164\pm0.015 \cdot X_{v} - 0.035\pm0.003 \cdot (V-I)$ & 0.022\\
     & Nov-2-96 & $I = i - 5.487\pm0.013 - 0.038\pm0.010 \cdot X_{i} + 0.004\pm0.002 \cdot (V-I)$ & 0.019\\
     &          &                                                                                 &      \\
LCO40& Dec-2-96 & $V = v - 3.456\pm0.010 - 0.150\pm0.009 \cdot X_{v} - 0.057\pm0.003 \cdot (V-I)$ & 0.021\\
     &          & $I = i - 3.848\pm0.009 - 0.104\pm0.008 \cdot X_{i} + 0.039\pm0.003 \cdot (V-I)$ & 0.018\\
     &          &                                                                                 &      \\
     & Dec-3-96 & $V = v - 3.490\pm0.009 - 0.118\pm0.006 \cdot X_{v} - 0.053\pm0.001 \cdot (V-I)$ & 0.015\\
     &          & $I = i - 3.909\pm0.008 - 0.048\pm0.006 \cdot X_{i} + 0.037\pm0.001 \cdot (V-I)$ & 0.012\\
     &          &                                                                                 &      \\
     & Dec-4-96 & $V = v - 3.437\pm0.013 - 0.173\pm0.009 \cdot X_{v} - 0.059\pm0.002 \cdot (V-I)$ & 0.023\\
     &          & $I = i - 3.805\pm0.019 - 0.121\pm0.014 \cdot X_{i} + 0.034\pm0.004 \cdot (V-I)$ & 0.020\\
     &          &                                                                                 &      \\
     & Dec-5-96 & $V = v - 3.458\pm0.016 - 0.153\pm0.012 \cdot X_{v} - 0.057\pm0.003 \cdot (V-I)$ & 0.016\\
     &          & $I = i - 3.792\pm0.023 - 0.129\pm0.016 \cdot X_{i} + 0.039\pm0.005 \cdot (V-I)$ & 0.023\\
     &          &                                                                                 &      \\
     & Dec-6-96 & $V = v - 3.438\pm0.011 - 0.155\pm0.008 \cdot X_{v} - 0.062\pm0.004 \cdot (V-I)$ & 0.013\\
     &          & $I = i - 3.885\pm0.018 - 0.079\pm0.014 \cdot X_{i} + 0.044\pm0.007 \cdot (V-I)$ & 0.020\\
     &          &                                                                                 &      \\
\tablenotetext{(a)}{The instrumental magnitudes were normalized to 1 second.}
\enddata
\end{deluxetable}

\begin{deluxetable}{rcccrcccrrclc}
\tablenum{3}
\tablecolumns{12}
\tablewidth{0pc}
\tablecaption{Photometric data and profile fit parameters of LSBD  galaxies in Dorado group \label{tab3}}
\tablehead{
\colhead{ID} &
\colhead{$\alpha$(2000)} &
\colhead{$\delta$(2000)} &
\colhead{V\tablenotemark{(a)}} &
\colhead{(V - I)\tablenotemark{(a)}} &
\colhead{$D_{26}$} &
\colhead{$\theta_{26}$} &
\colhead{$\mu_{0}$(V)} &
\colhead{h} &
\colhead{$r_{eff}$} &
\colhead{$\mu_{eff}$(V)} &
\colhead{Type} \\
\colhead{} &
\colhead{[$^{h \quad m \quad s}$]} &
\colhead{[$^{\circ \quad ' \quad ''}$]} &
\colhead{[mag]} &
\colhead{[mag]} &
\colhead{[$''$]} &
\colhead{[$''$]} &
\colhead{[mag/$\Box''$]} &
\colhead{[$''$]} &
\colhead{[$''$]} &
\colhead{[mag/$\Box''$]} &
\colhead{} \\ 
\colhead{(1)} &
\colhead{(2)} &
\colhead{(3)} &
\colhead{(4)} &
\colhead{(5)} &
\colhead{(6)} &
\colhead{(7)} &
\colhead{(8)} &
\colhead{(9)} &
\colhead{(10)} &
\colhead{(11)} &
\colhead{(12)} }
\startdata
  279-01 &  4 17 09.9 & -56 11 51 & 19.73 &  0.31 & 14.1 & 16.5 & 23.5 &  3.5 &  5.9 & 25.3 & \\
  377-01 &  4 12 23.7 & -56 11 12 & 18.50 &  0.95 & 16.5 & 18.6 & 22.7 &  3.0 &  5.1 & 24.5 & \\
  666-01 &  4 19 23.8 & -56 07 57 & 19.16 &  0.58 & 16.4 & 19.4 & 23.9 &  5.1 &  8.5 & 25.7 & \\
  779-01 &  4 14 02.3 & -56 07 38 & 19.74 &  2.08 & 14.4 & 15.6 & 23.6 &  3.6 &  5.9 & 25.4 & \\
  873-01 &  4 17 09.4 & -56 06 37 & 19.42 &  0.66 & 16.5 & 22.2 & 24.6 &  8.7 & 14.5 & 26.4 & dE \\
  874-01 &  4 18 15.4 & -56 06 20 & 20.00 &  1.43 & 14.5 & 16.9 & 24.1 &  4.8 &  8.0 & 25.9 & \\
  972-01 &  4 17 47.1 & -56 05 49 & 19.47 &  1.80 & 14.9 & 17.2 & 23.0 &  3.1 &  5.2 & 24.8 & \\
 1639-01 &  4 16 59.0 & -56 00 49 & 18.75 &  0.82 & 18.7 & 21.7 & 23.5 &  4.7 &  7.8 & 25.3 & dE\\
 1709-01 &  4 14 38.8 & -56 00 39 & 18.96 &  0.76 & 16.9 & 20.5 & 24.0 &  5.6 &  9.4 & 25.8 & \\
 1746-01 &  4 12 31.7 & -56 00 12 & 19.18 &  1.42 & 14.1 & 16.7 & 23.7 &  4.0 &  6.7 & 25.6 & \\
 1843-01 &  4 18 45.5 & -55 58 57 & 19.64 &  0.64 & 15.2 & 17.3 & 24.2 &  5.2 &  8.7 & 26.0 & \\
 2318-01 &  4 17 09.3 & -55 55 10 & 19.70 &  1.15 & 14.1 & 16.0 & 23.2 &  3.1 &  5.2 & 25.0 & \\
 2347-01 &  4 15 04.5 & -55 56 16 & 18.27 &  0.51 & 32.0 & 39.5 & 23.0 &  7.2 & 12.0 & 24.8 & dIrr\\
 2408-01 &  4 16 26.9 & -55 54 42 & 19.98 &  1.03 & 14.2 & 15.5 & 23.9 &  4.0 &  6.6 & 25.7 & \\
 2851-01 &  4 17 44.2 & -55 51 23 & 18.99 &  1.54 & 20.8 & 20.4 & 23.5 &  4.4 &  7.4 & 25.3 & \\
 2915-01 &  4 15 39.0 & -55 51 09 & 19.31 &  0.89 & 15.1 & 16.3 & 23.2 &  3.2 &  5.3 & 25.0 & \\
 2982-01 &  4 12 15.6 & -55 52 06 & 16.77 &  0.75 & 35.9 & 39.4 & 22.5 &  6.1 & 10.2 & 24.3 & dE\\
 3625-01 &  4 13 58.6 & -55 46 15 & 19.70 &  1.14 & 15.3 & 16.4 & 23.2 &  3.2 &  5.3 & 25.0 & \\
 3672-01 &  4 18 00.5 & -55 45 06 & 19.71 &  1.30 & 15.4 & 16.6 & 23.8 &  4.1 &  6.9 & 25.6 & \\
 3714-01 &  4 16 20.3 & -55 45 09 & 19.20 &  0.31 & 15.6 & 18.2 & 23.3 &  3.6 &  6.1 & 25.1 & \\
 3747-01 &  4 15 31.2 & -55 44 54 & 19.84 &  1.57 & 17.0 & 17.3 & 23.7 &  4.1 &  6.8 & 25.5 & \\
 3774-01 &  4 18 04.6 & -55 45 26 & 17.45 &  0.84 & 34.3 & 40.8 & 24.0 & 11.3 & 18.8 & 25.8 & dE\\
 3829-01 &  4 16 29.7 & -55 44 10 & 18.69 &  0.98 & 14.9 & 14.9 & 23.5 &  3.3 &  5.5 & 25.4 & \\
 4045-01 &  4 19 16.6 & -55 41 33 & 19.81 &  0.91 & 15.4 & 19.1 & 23.9 &  5.0 &  8.3 & 25.7 & \\
 4136-01 &  4 16 15.1 & -55 41 51 & 18.72 &  0.07 & 18.2 & 20.1 & 24.2 &  6.2 & 10.3 & 26.0 & \\
 4302-01 &  4 14 09.0 & -55 42 25 & 15.74 &  0.85 & 68.9 & 69.6 & 22.7 & 11.3 & 18.9 & 24.5 & dE\\
 4429-01 &  4 19 29.4 & -55 39 00 & 18.38 &  0.85 & 22.4 & 26.8 & 23.9 &  7.0 & 11.7 & 25.7 & dE\\
 4488-01 &  4 13 24.8 & -55 38 53 & 19.62 &  0.88 & 15.4 & 17.4 & 24.2 &  5.2 &  8.8 & 26.0 & dE\\
 4536-01 &  4 17 11.5 & -55 38 10 & 19.66 &  2.07 & 15.8 & 16.7 & 23.0 &  3.0 &  5.1 & 24.8 & \\
 4598-01 &  4 14 52.7 & -55 38 09 & 19.44 &  0.85 & 15.9 & 15.5 & 23.6 &  3.5 &  5.8 & 25.4 & \\
 4814-01 &  4 14 40.5 & -55 36 02 & 19.11 &  0.99 & 14.1 & 17.0 & 23.5 &  3.7 &  6.1 & 25.3 & \\
 5012-01 &  4 16 00.4 & -55 34 06 & 19.04 &  1.14 & 15.1 & 14.4 & 24.2 &  4.4 &  7.4 & 26.1 & \\
 5192-01 &  4 16 22.9 & -55 32 29 & 19.96 &  1.03 & 15.3 & 14.8 & 23.5 &  3.2 &  5.3 & 25.3 & \\
 5242-01 &  4 18 08.6 & -55 32 08 & 19.11 &  0.93 & 16.7 & 17.8 & 23.0 &  3.2 &  5.3 & 24.8 & \\
 5622-01 &  4 14 55.0 & -55 29 16 & 19.06 &  1.01 & 16.5 & 17.5 & 23.0 &  3.1 &  5.2 & 24.8 & \\
 5894-01 &  4 16 09.1 & -55 27 33 & 19.37 &  0.41 & 18.2 & 22.5 & 23.9 &  6.0 & 10.0 & 25.8 & \\
 5923-01 &  4 18 37.1 & -55 26 52 & 19.54 &  1.21 & 14.3 & 15.5 & 23.7 &  3.6 &  6.0 & 25.5 & \\
 6672-01 &  4 16 48.3 & -55 16 05 & 19.13 &  1.23 & 16.1 & 19.3 & 23.1 &  3.6 &  6.0 & 24.9 & \\
 6740-01 &  4 19 26.9 & -55 16 11 & 18.85 &  0.12 & 20.8 & 21.0 & 22.6 &  3.4 &  5.6 & 24.4 & \\
 6854-01 &  4 14 09.8 & -55 19 22 & 19.34 &  1.25 & 16.9 & 17.3 & 23.5 &  3.7 &  6.2 & 25.3 & \\
 6942-01 &  4 14 44.6 & -55 18 25 & 19.89 &  1.52 & 14.5 & 17.2 & 24.0 &  4.8 &  8.0 & 25.9 & \\
 7022-01 &  4 12 53.3 & -55 19 30 & 19.29 &  0.24 & 14.9 & 18.7 & 23.7 &  4.4 &  7.4 & 25.5 & \\
 7328-01 &  4 12 56.6 & -55 22 34 & 18.99 &  0.71 & 18.9 & 20.5 & 22.7 &  3.4 &  5.7 & 24.5 & \\
 7330-01 &  4 16 07.0 & -55 22 10 & 18.26 &  0.25 & 25.9 & 28.9 & 24.4 &  9.9 & 16.6 & 26.2 & dE\\
 7392-01 &  4 12 18.4 & -55 22 54 & 18.64 &  1.37 & 24.4 & 26.1 & 22.8 &  4.4 &  7.3 & 24.6 & \\
 7493-01 &  4 14 20.7 & -55 23 36 & 19.79 &  1.36 & 15.2 & 16.4 & 23.7 &  3.8 &  6.4 & 25.5 & \\
 7547-01 &  4 16 18.0 & -55 23 29 & 19.23 &  0.30 & 14.6 & 17.1 & 23.8 &  4.3 &  7.2 & 25.6 & \\
 7731-01 &  4 17 13.0 & -55 24 59 & 19.64 &  0.95 & 16.5 & 19.4 & 23.6 &  4.4 &  7.4 & 25.4 & \\
 7778-01 &  4 17 59.0 & -55 24 44 & 18.95 &  0.46 & 24.2 & 25.5 & 23.1 &  4.9 &  8.1 & 25.0 & \\
 7783-01 &  4 18 56.9 & -55 25 02 & 19.41 &  1.40 & 16.4 & 19.5 & 23.0 &  3.5 &  5.8 & 24.8 & \\
 7952-01 &  4 17 16.6 & -55 25 03 & 19.53 &  1.60 & 14.9 & 17.2 & 23.4 &  3.7 &  6.1 & 25.3 & \\
 1064-03 &  4 15 50.1 & -54 40 05 & 18.34 &  1.10 & 24.3 & 27.9 & 23.5 &  6.2 & 10.3 & 25.4 & dE\\
   88-05 &  4 13 15.9 & -56 44 08 & 19.74 &  1.07 & 14.0 & 15.8 & 23.8 &  3.8 &  6.4 & 25.6 & dE\\
 1977-05 &  4 11 03.3 & -56 24 27 & 19.13 &  0.82 & 14.7 & 16.2 & 23.2 &  3.1 &  5.2 & 25.0 & dE\\
 2674-06 &  4 10 35.3 & -57 10 54 & 19.84 & -0.23 & 15.4 & 17.5 & 23.1 &  3.3 &  5.5 & 25.0 & dIrr\tablenotemark{(b)}\\
 1374-09 &  4 27 13.2 & -55 38 52 & 18.67 &  0.85 & 18.4 & 21.9 & 22.8 &  3.8 &  6.3 & 24.7 & dE\\
 3330-10 &  4 03 04.9 & -55 18 53 & 18.46 &  2.26 & 21.1 & 20.5 & 22.7 &  3.4 &  5.7 & 24.5 & dIrr\tablenotemark{(c)}\\
 1116-11 &  4 23 41.1 & -56 45 03 & 18.17 &  0.94 & 18.5 & 23.3 & 22.7 &  3.8 &  6.4 & 24.5 & dIrr\\
 1649-11 &  4 22 33.1 & -56 41 09 & 19.00 &  1.15 & 19.3 & 21.3 & 24.2 &  6.4 & 10.7 & 26.0 & dE\\
 3041-11 &  4 22 52.9 & -56 37 27 & 17.89 &  0.72 & 33.8 & 37.9 & 23.3 &  7.6 & 12.7 & 25.1 & dIrr\tablenotemark{(b)}\\
3854-d-02&  4 15 17.0 & -54 58 57 & 20.08 &  1.63 & 14.8\tablenotemark{(e)} & 16.8\tablenotemark{(e)} & 22.9 &  2.6 &  4.3 & 24.7 & dIrr\tablenotemark{(b)} \\
 802-d-07&  4 23 19.2 & -55 46 45 & 18.85 &  1.00 & 15.6\tablenotemark{(e)} & 17.0\tablenotemark{(e)} & 22.5 &  2.3 &  3.9 & 24.4 & dE \\
1961-d-07&  4 21 05.3 & -55 40 10 & 18.95 &  1.47 & 18.4\tablenotemark{(e)} & 16.4\tablenotemark{(e)} & 22.6 &  2.3 &  3.9 & 24.5 & dE \\
2029-d-08&  4 10 33.5 & -55 30 15 & 19.57 &  0.80 & 14.7\tablenotemark{(e)} & 15.5\tablenotemark{(e)} & 24.6 &  3.8 &  6.4 & 26.4 & dE \\
3714-d-10&  4 02 07.8 & -55 19 24 & 19.49 &  0.95 & 14.3\tablenotemark{(e)} & 16.0\tablenotemark{(e)} & 23.5 &  2.9 &  4.8 & 25.3 & dE \\
L01\tablenotemark{(f)} &  4 12 25.5 & -56 05 31 & 18.50 &\nodata& 29.2 & 30.4 & 24.8 & 13.3 & 22.2 & 26.6 & dE\\
L02\tablenotemark{(f)} &  4 14 56.2 & -55 31 22 & 17.71 &\nodata& 38.4 & 39.1 & 24.9 & 19.6 & 32.7 & 26.7 & dE\tablenotemark{(d)}\\
L03\tablenotemark{(f)} &  4 14 37.4 & -55 25 27 & 18.97 &\nodata& 22.7 & 29.9 & 24.7 &  7.5 & 12.5 & 26.5 & dE\\
L04\tablenotemark{(f)} &  4 14 44.2 & -55 24 54 & 19.12 &\nodata& 23.2 & 25.7 & 24.9 & 12.9 & 21.5 & 26.7 & dE\\
\tablenotetext{(a)}{Not corrected according to the simulation in sec. 2.6 and 
not corrected by interstellar absorption.}
\tablenotetext{(b)}{Shows knots of star formation regions.}
\tablenotetext{(c)}{Red irregular galaxy.}
\tablenotetext{(d)}{Extended irregular low-surface brightness cloud located $\sim10^{'}$ to the NW of the NGC1549 galaxy.}
\tablenotetext{(e)}{Limiting isophotal diameter and diameter given by relation (4) at the surface brightness of 26.5 V mag/arcsec$^{2}$.}
\tablenotetext{(f)}{These galaxies were not detected in the I filter (see section 2.7).}
\enddata
\end{deluxetable}

\begin{deluxetable}{ccccc}
\tablenum{4}
\tablecolumns{5}
\tablewidth{0pc}
\tablecaption{Completeness fraction for point sources \label{tab4}}
\tablehead{
\colhead{Magnitude} & 
\colhead{f(V)\tablenotemark{(a)}} & 
\colhead{$\sigma_{f(V)}$\tablenotemark{(b)}} & 
\colhead{f(V)\tablenotemark{(a)}} & 
\colhead{$\sigma_{f(V)}$\tablenotemark{(b)}} \\
\colhead{(bin)} & 
\colhead{[CTIO]} & 
\colhead{[CTIO]} & 
\colhead{[LCO]} & 
\colhead{[LCO]} \\
\colhead{(1)} &
\colhead{(2)} &
\colhead{(3)} &
\colhead{(4)} &
\colhead{(5)} }
\startdata
16.0 - 16.5 & 0.99 & 0.04 & 1.00 & 0.01 \\
16.5 - 17.0 & 0.98 & 0.04 & 1.00 & 0.01 \\
17.0 - 17.5 & 0.98 & 0.03 & 1.00 & 0.01 \\
17.5 - 18.0 & 0.96 & 0.03 & 1.00 & 0.01 \\
18.0 - 18.5 & 0.96 & 0.05 & 1.00 & 0.01 \\
18.5 - 19.0 & 0.96 & 0.05 & 1.00 & 0.01 \\
19.0 - 19.5 & 0.93 & 0.05 & 0.99 & 0.03 \\
19.5 - 20.0 & 0.91 & 0.06 & 0.99 & 0.03 \\
20.0 - 20.5 & 0.88 & 0.07 & 0.99 & 0.04 \\
20.5 - 21.0 & 0.45 & 0.13 & 0.97 & 0.04 \\
21.0 - 21.5 &\nodata&\nodata& 0.97 & 0.03 \\
21.5 - 22.0 &\nodata&\nodata& 0.95 & 0.04 \\
22.0 - 22.5 &\nodata&\nodata& 0.92 & 0.06 \\
22.5 - 23.0 &\nodata&\nodata& 0.92 & 0.06 \\
23.0 - 23.5 &\nodata&\nodata& 0.89 & 0.08 \\
23.5 - 24.0 &\nodata&\nodata& 0.66 & 0.10 \\
\tablenotetext{(a)}{f(V): detected fraction in the V band}
\tablenotetext{(b)}{$\sigma_{f(V)}$: one standard deviation in f(V)}
\enddata
\end{deluxetable}

\begin{deluxetable}{ccccc}
\tablenum{5}
\tablecolumns{5}
\tablewidth{0pc}
\tablecaption{Average completeness fraction for LSB disk galaxies \label{tab5}}
\tablehead{
\colhead{Magnitude} & 
\colhead{f(V)\tablenotemark{(a)}} & 
\colhead{$\sigma_{f(V)}$\tablenotemark{(b)}} &
\colhead{f(V)\tablenotemark{(a)}} & 
\colhead{$\sigma_{f(V)}$\tablenotemark{(b)}} \\
\colhead{(bin)} & 
\colhead{[CTIO]} & 
\colhead{[CTIO]} & 
\colhead{[LCO]} & 
\colhead{[LCO]} \\
\colhead{(1)} &
\colhead{(2)} &
\colhead{(3)} &
\colhead{(4)} &
\colhead{(5)} }
\startdata
16.0 - 16.5 & 0.77 & 0.04 & 0.75 & 0.08 \\
16.5 - 17.0 & 0.83 & 0.01 & 0.83 & 0.04 \\
17.0 - 17.5 & 0.83 & 0.01 & 0.82 & 0.09 \\
17.5 - 18.0 & 0.84 & 0.06 & 0.85 & 0.07 \\
18.0 - 18.5 & 0.82 & 0.07 & 0.91 & 0.07 \\
18.5 - 19.0 & 0.71 & 0.06 & 0.86 & 0.04 \\
19.0 - 19.5 & 0.73 & 0.06 & 0.87 & 0.03 \\
19.5 - 20.0 & 0.53 & 0.10 & 0.84 & 0.06 \\
20.0 - 20.5 &\nodata&\nodata& 0.78 & 0.07 \\
20.5 - 21.0 &\nodata&\nodata& 0.60 & 0.27 \\  
\tablenotetext{(a)}{f(V): detected fraction in the V band}
\tablenotetext{(b)}{$\sigma_{f(V)}$: one standard deviation in f(V)}
\enddata
\end{deluxetable}

\begin{deluxetable}{cccrcrcrc}
\tablenum{6}
\tablecolumns{9}
\tablewidth{0pc}
\tablecaption{Monte Carlo simulations: photometric and scale length errors \label{tab6}}
\tablehead{
\colhead{ Session } &
\colhead{ CSB interval\tablenotemark{(a)} } &
\colhead{ V$_{inp}$} &
\colhead{ $\Delta (V)$\tablenotemark{(b)} } &
\colhead{ $\sigma_{\Delta(V)}$ \tablenotemark{(c)} } &
\colhead{ $\Delta(\mu_{0}) $\tablenotemark{(d)} } &
\colhead{ $\sigma_{\Delta(\mu_{0})} $\tablenotemark{(e)} } &
\colhead{ $\Delta h/h$\tablenotemark{(f)} } &
\colhead{ $\sigma_{\Delta h/h}$\tablenotemark{(g)}} \\
\colhead{(1)} &
\colhead{(2)} &
\colhead{(3)} &
\colhead{(4)} &
\colhead{(5)} &
\colhead{(6)} &
\colhead{(7)} &
\colhead{(8)} &
\colhead{(9)}}
\startdata
CTIO & 22.5 - 23.5 & 16.25 &  -0.08 & 0.05 &  0.09 & 0.09 & -0.06 & 0.06 \\
     &             & 16.75 &  -0.07 & 0.06 &  0.05 & 0.10 & -0.08 & 0.06 \\
     &             & 17.25 &  -0.07 & 0.08 & -0.02 & 0.11 & -0.13 & 0.08 \\
     &             & 17.75 &  -0.07 & 0.07 & -0.08 & 0.13 & -0.18 & 0.09 \\
     & 23.5 - 24.5 & 16.25 &  -0.17 & 0.09 &  0.20 & 0.09 & -0.01 & 0.07 \\
     &             & 16.75 &  -0.18 & 0.08 &  0.18 & 0.11 & -0.02 & 0.08 \\
     &             & 17.25 &  -0.14 & 0.10 &  0.15 & 0.13 & -0.04 & 0.09 \\
     &             & 17.75 &  -0.13 & 0.12 &  0.11 & 0.14 & -0.10 & 0.15 \\
     &             & 18.25 &  -0.13 & 0.11 &  0.07 & 0.18 & -0.10 & 0.14 \\
     &             & 18.75 &  -0.13 & 0.13 &  0.02 & 0.19 & -0.16 & 0.17 \\
     & 24.5 - 25.5\tablenotemark{(h)} & 17.75 &  -0.27 & 0.23 & -0.56 & 0.11 & -0.20 & 0.25 \\
     &             & 18.25 &  -0.29 & 0.20 & -0.93 & 0.16 & -0.20 & 0.11 \\
     &             & 18.75 &  -0.24 & 0.21 & -1.23 & 0.13 & -0.35 & 0.33 \\
     &             & 19.25 &  -0.24 & 0.19 & -1.49 & 0.17 & -0.40 & 0.31 \\
     &             & 19.75 &   0.10 & 0.51 & -1.64 & 0.21 & -0.43 & 0.41 \\
LCO40& 22.5 - 23.5 & 16.25 &  -0.09 & 0.05 &  0.08 & 0.08 & -0.03 & 0.03 \\
     &             & 16.75 &  -0.07 & 0.04 &  0.04 & 0.09 & -0.05 & 0.04 \\
     &             & 17.25 &  -0.06 & 0.04 &  0.03 & 0.09 & -0.06 & 0.04 \\
     &             & 17.75 &  -0.04 & 0.05 &  0.02 & 0.09 & -0.07 & 0.05 \\
     &             & 18.25 &  -0.05 & 0.05 & -0.02 & 0.09 & -0.09 & 0.05 \\
     &             & 18.75 &  -0.04 & 0.05 & -0.05 & 0.11 & -0.12 & 0.06 \\
     & 23.5 - 24.5 & 16.25 &  -0.13 & 0.11 &  0.14 & 0.09 &  0.02 & 0.05 \\
     &             & 16.75 &  -0.11 & 0.09 &  0.13 & 0.09 & -0.01 & 0.05 \\
     &             & 17.25 &  -0.15 & 0.08 &  0.13 & 0.10 & -0.03 & 0.06 \\
     &             & 17.75 &  -0.13 & 0.08 &  0.13 & 0.10 & -0.05 & 0.06 \\
     &             & 18.25 &  -0.10 & 0.08 &  0.09 & 0.08 & -0.07 & 0.06 \\
     &             & 18.75 &  -0.09 & 0.06 &  0.10 & 0.10 & -0.08 & 0.08 \\
     &             & 19.25 &  -0.07 & 0.09 &  0.04 & 0.11 & -0.10 & 0.09 \\
     &             & 19.75 &  -0.08 & 0.09 &  0.01 & 0.13 & -0.14 & 0.08 \\
     & 24.5 - 25.5 & 16.75 &  -0.27 & 0.18 &  0.26 & 0.11 &  0.16 & 0.11 \\
     &             & 17.25 &  -0.21 & 0.17 &  0.29 & 0.10 &  0.10 & 0.10 \\
     &             & 17.75 &  -0.22 & 0.18 &  0.27 & 0.11 &  0.07 & 0.11 \\
     &             & 18.25 &  -0.18 & 0.19 &  0.27 & 0.11 &  0.05 & 0.10 \\
     &             & 18.75 &  -0.21 & 0.15 &  0.28 & 0.12 &  0.01 & 0.10 \\
     &             & 19.25 &  -0.19 & 0.16 &  0.27 & 0.12 &  0.00 & 0.10 \\
     &             & 19.75 &  -0.18 & 0.16 &  0.19 & 0.14 & -0.02 & 0.10 \\
     &             & 20.25 &  -0.14 & 0.16 &  0.14 & 0.16 & -0.14 & 0.14 \\
     &             & 20.75 &  -0.01 & 0.17 &  0.11 & 0.22 & -0.34 & 0.14 \\
     & 25.5 - 26.5\tablenotemark{(h)} & 17.75 &  -0.29 & 0.24 & -0.22 & 0.10 &  0.03 & 0.15 \\
     &             & 18.25 &  -0.16 & 0.20 & -0.19 & 0.14 &  0.03 & 0.13 \\
     &             & 18.75 &  -0.15 & 0.19 & -0.26 & 0.12 & -0.06 & 0.16 \\
     &             & 19.25 &  -0.12 & 0.14 & -0.36 & 0.12 & -0.07 & 0.12 \\
     &             & 19.75 &  -0.16 & 0.18 & -0.33 & 0.15 & -0.12 & 0.21 \\
     &             & 20.25 &  -0.12 & 0.15 & -0.47 & 0.14 & -0.34 & 0.37 \\
     &             & 20.75 &  -0.33 & 0.41 & -0.61 & 0.20 & -0.25 & 0.14 \\
\tablenotetext{(a)}{Central surface brightness interval (mag/arcsec$^{2}$).}
\tablenotetext{(b)}{$\Delta(V)=V_{inp}-V_{out}$: difference bewteen input and
output magnitudes.}
\tablenotetext{(c)}{$\sigma_{\Delta(V)}$: one standard deviation in
$\Delta(V)$.}
\tablenotetext{(d)}{$\Delta(\mu_{0})=\mu^{inp}_{0}-\mu^{out}_{0}$: differences
between input and output central surface brightnesses (mag/arcsec$^{2}$).}
\tablenotetext{(e)}{$\sigma_{\Delta(\mu_{0})}$: one standard deviation in
$\Delta(\mu_{0})$ (mag/arcsec$^{2}$).}
\tablenotetext{(f)}{$\Delta h/h = (h_{inp} - h_{out})/h_{inp}$: differences 
between input and output scale length in percentage.}
\tablenotetext{(g)}{$\sigma_{\Delta h/h}$: one standard deviation in $\Delta h/h$.}
\tablenotetext{(h)}{The results showed in the last bins of surface brightnesses for the CTIO and LCO40 data were obtained using the smoothing method described in section 2.7.}
\enddata
\end{deluxetable}

\end{document}